\documentclass[reprint,amsmath,amssymb,aps,prb,superscriptaddress,twocolumn]{revtex4-2}
\usepackage{graphicx}
\usepackage{graphics}
\usepackage{dcolumn}
\usepackage{bm}
\usepackage{amsfonts}
\usepackage{amssymb}
\usepackage{xcolor}
\usepackage{multirow}
\usepackage{mathtools}
\usepackage[tight]{subfigure}
\usepackage{relsize}
\usepackage{float}

\definecolor{green}{rgb}{0,0.6,0.1}

\begin{document}

\title{{\it Ab initio} prediction of high temperature $d_{x^2-y^2}$-wave superconductivity in infinite-layer nickelates}
\title{{\it Ab initio} prediction of $d_{x^2-y^2}$-wave superconductivity in infinite-layer nickelates}


\author{Guang-Yu Guo}
\affiliation{Department of Physics, National Taiwan University, Taipei 10617, Taiwan\looseness=-1}
\affiliation{Physics Division, National Center for Theoretical Sciences, Taipei 10617, Taiwan\looseness=-1}

\author{Ren-Guo Guo}
\affiliation{Department of Physics, National Taiwan University, Taipei 10617, Taiwan\looseness=-1}

\author{Yun-Chen Liao}
\affiliation{Department of Physics, National Taiwan University, Taipei 10617, Taiwan\looseness=-1}

\author{Yang-hao Chan}
\affiliation{Institute of Atomic and Molecular Sciences, Academia Sinica, Taipei 10617, Taiwan\looseness=-1}
\affiliation{Physics Division, National Center for Theoretical Sciences, Taipei 10617, Taiwan\looseness=-1}

\date{\today}

\begin{abstract}
Infinite-layer nickelates have recently emerged as a new family of potential 
unconventional high critical temperature ($T_c$) superconductors.
However, fundamental questions such as their superconducting (SC) pairing mechanism and gap symmetry remain 
under intense debate. Here we present a fully {\it ab initio} theoretical study on the 
SC properties of optimally doped nickelates $Re$$_{0.8}$Sr$_{0.2}$NiO$_2$ ($Re=$ La, Pr, Nd), 
based on the density functional theory for superconductors calculations with electron-phonon 
coupling (EPC), screened Coulomb repulsion and spin fluctuation (SF) interaction 
treated on an equal footing. We find that $Re_{0.8}$Sr$_{0.2}$NiO$_2$ are two-band superconductors 
with sign reversal $d_{x^2-y^2}(\pm)$-wave gap functions on the different Fermi surface (FS) pockets.
Interestingly, when the SF interaction is turned off, $T_c$ becomes negligibly small ($\sim$0.01 K), 
thus demonstrating that the superconductivity in $Re_{0.8}$Sr$_{0.2}$NiO$_2$ is driven by SF interaction.
Moreover, our {\it ab initio} calculations reveal that the SF interaction is an order of magnitude stronger
than both EPC and Coulomb repulsion on the large quasi-two-dimensional FS pocket around the Brillouin
zone (BZ) center, thus leading to the SF-mediated pairing mechanism, although the EPC dominates on 
the small three-dimensional electron FS pockets at the BZ corners. 
The emergence of nodal $d_{x^2-y^2}(\pm)$-wave gap structure is traced to the pronounced peaks 
in the Lindhard response function at the BZ corners.
Our calculated FS, SC critical temperature, nodal gap structure and SC quasiparticle 
density of states are consistent with most available experiments. 
Furthermore, predicted unconventional SC properties such as scanning tunneling spectra 
of La$_{0.8}$Sr$_{0.2}$NiO$_2$ and Pr$_{0.8}$Sr$_{0.2}$NiO$_2$ are ready for
immediate experimental verifications.
\end{abstract}

\maketitle
\section{INTRODUCTION}
\label{sec:introduction}
High critical temperature ($T_c$) superconductivity was discovered in copper oxides (cuprates) 
in late 1980s.~\cite{Bednorz1986,Wu1987}
With the $T_c$ exceeding the boiling point of liquid nitrogen (77 K)~\cite{Wu1987,Schilling1993}, 
the cuprates promised spectacular applications such as electrical networks 
with no loss of electrical energy, and have triggered intensive investigations ever since.
The intensive studies in the past four decades have ruled out the conventional $s$-wave phonon-mediated 
Bardeen-Cooper-Schrieffer (BCS) mechanism~\cite{Bardeen1957} and have established, at least qualitatively, 
unconventional $d$-wave pairing symmetry in the cuprates~\cite{Scalapino1999,Tsuei2000,Hashimoto2014,Keimer2015}. 
However, fundamental questions such as the origin of the high $T_c$ superconductivity
and mechanism of electron pairing remain 
unanswered.~\cite{Tsuei2000,Hashimoto2014,Keimer2015,Luo2023,Wen2025}
Meanwhile, due to nickel's proximity to copper in the periodic table, 
nickel compounds with similar crystalline and electronic 
structure have been studied with the aim of finding additional high $T_c$ superconductors
and better understanding the superconductivity in the 
cuprates (see, e.g.,~\cite{Guo1988,Anisimov1999,Lee2004,Hansmann2009}). 
This endeavor culminated in the recent discovery of superconductivity in hole-doped 
infinite-layer nickelate Nd$_{1-x}$Sr$_x$NiO$_2$ ($x= 0.1–0.3$) with rather high $T_c$.~\cite{Li2019}
At present, several nickelate superconductors have been 
discovered~\cite{Li2019,Zeng2020,Osada2020,Li2020,Osada2021,Lee2023,Osada2023,Chow2025},
including Nd$_{1-x}$Sr$_x$NiO$_2$, Pr$_{1-x}$Sr$_x$NiO$_2$ and La$_{1-x}$Sr$_x$NiO$_2$.

The superconducting (SC) pairing interaction and gap symmetry in infinite-layer nickelates 
are under heated debate. Early {\it ab initio} density functional theory (DFT) calculations~\cite{Nomura2019} 
showed that the electron-phonon coupling (EPC) in NdNiO$_2$ is too weak to account for the observed $T_c$
of $\sim$20 K in Nd$_{1-x}$Sr$_x$NiO$_2$.
Infinite-layer nickelates and cuprates share a similar transition-metal-oxygen 
square planar lattice, with Ni$^{1+}$ and Cu$^{2+}$ in the same 3$d^9$ valence 
configuration.~\cite{Lee2004,Li2019,Botana2020,Nomura2022}.
This similarity in the crystal structure and electronic structure 
has motivated proposals of a compelling scenario of unconventional $d$-wave superconductivity 
in the nickelates, supported by a random phase approximation (RPA)~\cite{Wu2020} 
and a fluctuation exchange approximation~\cite{Sakakibara2020} calculation based on the tight-binding model.
Indeed, recent superfluid density experiments on La$_{0.8}$Sr$_{0.2}$NiO$_2$
and Nd$_{0.8}$Sr$_{0.2}$NiO$_2$~\cite{Harvey2025} showed a quadratic temperature dependence, 
indicating nodal superconductivity. However, a scanning tunneling spectroscopy (STS) experiment 
on Nd$_{1-x}$Sr$_x$NiO$_2$ ($x= 0.12-0.25$)~\cite{Gu2020} observed SC gap structures 
consistent with both $d$-wave and $s$-wave depending 
on the tip position during the measurements. Furthermore, another superfluid density experiment~\cite{Chow2022} 
suggested that the SC order parameter in Nd$_{0.8}$Sr$_{0.2}$NiO$_2$ and La$_{0.8}$Sr$_{0.2}$NiO$_2$ 
is beyond a single $d$-wave gap.
A recent {\it ab initio} study~\cite{Li2024}, based on the GW and GW perturbation theory (GWPT) 
calculations, predicted a phonon-mediated two-gap $s$-wave superconductivity in Nd$_{1-x}$Sr$_x$NiO$_2$
with the calculated $T_c$ agreeing with experiments~\cite{Lee2023}.
The {\it ab initio} GW approach has achieved much success in describing quasiparticle properties of
many materials including the EPC.~\cite{Li2024} 
However, the role of spin fluctuation (SF) interaction, which presumably is the main driving force for 
the $d$-wave pairing in cuprates~\cite{Scalapino1999,Tsuei2000,Keimer2015}, 
has not been investigated.

Here we apply fully {\it ab initio} density functional theory for superconductors (SCDFT)~\cite{Oliveira1988}
to study optimally Sr-doped infinite-layer
nickelates $Re_{0.8}$Sr$_{0.2}$NiO$_2$ ($Re$SrNiO) ($Re =$ La, Pr, Nd).
SCDFT extends DFT to account for gauge symmetry breaking in superconductors.
It was later extended to multi-component DFT~\cite{Lueders2005,Marques2005},
incorporating various many-body interactions. In the current formalism of the SCDFT~\cite{Kawamura2020},
EPC, dynamically screened electron-electron (e-e) Coulomb
repulsion and SF-mediated pairing interaction are all treated in a first-principles manner
(see, e.g., \cite{Lueders2005,Marques2005,Akashi2013,Essenberger2014,Kawamura2020} and references therein).
Thus, the SCDFT provides a fully {\it ab initio} framework for both conventional
phonon-mediated superconductors such as MgB$_2$~\cite{Floris2005} and for unconventional pairing mechanism
driven by e-e interactions such as spin-fluctuations~\cite{Essenberger2014,Kawamura2020}
and plasmon oscillations~\cite{Akashi2013,Kawamura2020}. See the next two sections for more information
on the theory and computational details.
Remarkably, our {\it ab initio} SCDFT calculations reveal that $Re_{0.8}$Sr$_{0.2}$NiO$_2$ are 
unconventional $d_{x^2-y^2}(\pm)$-wave superconductors with a $B_{1g}$ order parameter. 
Their SC gap structure consists of $d_{x^2-y^2}$-waves 
with opposite signs on the two disconnected Fermi surface (FS) pockets. The calculated $T_c$ values agree 
well with the experiments for all three considered nickelates.
Moreover, we demonstrate that the $d_{x^2-y^2}$-wave superconductivity in $Re$SrNiO is driven 
by the antiferromagnetic SFs with wavevectors near ($\pi/a,\pi/a$) in these compounds. \\ 

The rest of this paper is organized as follows. In the next section, we provide a brief description of 
the {\it ab initio} SCDFT, while the crystal structure of $Re_{0.8}$Sr$_{0.2}$NiO$_2$ 
and the computational details are given in Sec.~\ref{sec:details}.
In Sec.~\ref{sec:Electronic_Structure}, we report the calculated electronic structure
and Fermi surface of $Re_{0.8}$Sr$_{0.2}$NiO$_2$. 
In Sec. ~\ref{sec:SC}, the calculated SC properties are presented, 
including temperature-dependent SC gap values and momentum ${\bf{k}}$-resolved SC gap function on the FS.
Also in this section, the symmetry and structure of the obtained gap functions are analyzed in terms 
of the point group of the crystal structure of the nickelates.
In Sec. ~\ref{sec:origin}, the origin and mechanism of the unconventional $d$-wave
superconductivity uncovered in this work are revealed by our "computer experiments" and 
explained with the peaks in the calculated Lindhard response functions.  
In Sec. VII, our predictions are compared with available experiments and it is concluded that
our calculated FS, $T_c$, nodal gap structure and quasiparticle density of states are consistent with 
most available experiments.  Further experiments to clarify the remaining questions are proposed.
In Appendix A, we provide irreducible representations and basis functions
for point group $D_{4h}$, which are needed to understand the symmetry and structure of
the SC gap functions of the nickelates. In Appendix B, the calculated phonon dispersion,
EPC and phonon-mediated superconductivity are reported. 
Finally, in Appendix C, band- and momentum-dependent EPC strength, screened Coulomb repulsion,
and SF interaction on the FS are presented in order to
gain in-depth insight into the SC pairing mechanism and the SC gap structure in $Re_{0.8}$Sr$_{0.2}$NiO$_2$.

\section{Density functional theory for superconductors}
\label{sec:SCDFT}
In the SCDFT, starting with the normal state properties
from conventional DFT calculations as inputs, one solves
the BCS-like superconducting (SC) gap equation (see, e.g., ~\cite{Lueders2005,Marques2005,Kawamura2020})
\begin{align}
\Delta_{n\bf{k}} = - \frac{1}{2}\mathlarger{\sum}_{m{\bf k}'}\frac{K_{n{\bf k}m{\bf k}'}(\xi_{n{\bf k}},\xi_{m{\bf k}'})}{1+Z_{n{\bf k}}(\xi_{n{\bf k}})}\frac{\mathrm{tanh}[(\beta/2)E_{m{\bf k}'}]}{E_{m{\bf k}'}}\Delta_{m{\bf k}'},
\label{eq:gap-eq}
\end{align}
where $\beta$ is the inverse temperature ($1/k_BT$),  $\Delta_{n{\bf k}}$ is the gap function, $n$ and $\bf k$
denote the band index and the crystal momentum, respectively.
Also, $E_{n\bf{k}} = \sqrt{\xi_{n{\bf k}}^2+|\Delta_{n{\bf k}}|^2}$
and $\xi_{n{\bf k}} = \varepsilon_{n{\bf k}} - \mu$ which is the DFT eigen-energy ($\varepsilon_{n{\bf k}}$)
measured from the chemical potential $\mu$.
The integration kernels $K_{n{\bf k}m{\bf k}'}(\xi_{n{\bf k}},\xi_{m{\bf k}'})$ include the superconducting-pair
breaking and creating interactions and comprises the EPC, the screened e-e Coulomb
repulsion, and the SF kernel,
i.e., $K_{n{\bf k}m{\bf k}'}(\xi_{n{\bf k}},\xi_{m{\bf k}'}) = K^{ep}_{n{\bf k}m{\bf k}'}(\xi_{n{\bf k}},\xi_{m{\bf k}'})
+ K^{ee}_{n{\bf k}m{\bf k}'}(\xi_{n{\bf k}},\xi_{m{\bf k}'}) + K^{SF}_{n{\bf k}m{\bf k}'}(\xi_{n{\bf k}},\xi_{n'{\bf k}'})$.
The renormalization $Z_{n{\bf k}}(\xi_{n{\bf k}})$ contains only the EPC and SF terms, i.e.,
$Z_{n{\bf k}}(\xi_{n{\bf k}}) = Z_{n{\bf k}}^{ep}(\xi_{n{\bf k}}) + Z_{n{\bf k}}^{SF}(\xi_{n{\bf k}})$,
because the screened e-e Coulomb repulsion is already included in the DFT eigenvalues $\xi_{n{\bf k}}$.
The expressions for kernels $K_{n{\bf k}m{\bf k}'}(\xi_{n{\bf k}},\xi_{m{\bf k}'})$ and renormalization
$Z_{n{\bf k}}(\xi_{n{\bf k}})$ have already been given in, e.g., ~\cite{Lueders2005,Marques2005,Essenberger2014,Kawamura2020}.

The central problem in the SCDFT calculations is to solve the gap equation [Eq.~(\ref{eq:gap-eq})]
self-consistently. This requires not only the prior {\it ab initio} DFT calculation of the normal
state electronic energy bands ($\varepsilon_{n{\bf k}}$) and phonon dispersion
but also the prior {\it ab initio} calculation of EPC matrix elements ($g_{mn,\nu}(\bf{k},\bf{q})$)
using the density functional perturbation theory (DFPT)~\cite{Baroni2001} as well as the charge
and spin susceptibility using the DFT within either the random phase approximation (RPA)~\cite{Gell-Mann1957,Kawamura2020}
or adiabatic local density approximation (ALDA)~\cite{Zangwill1980,Tsutsumi2020}.
We find that the SC properties of $Re_{0.8}$Sr$_{0.2}$NiO$_2$ calculated with the RPA and ALDA do not differ significantly.
Thus, in the present paper, we present the calculated SC properties using the ALDA only. \\

\section{Crystal structure and computational details}
\label{sec:details}
\begin{figure}[!htb]
\centering
\includegraphics[width=0.85\columnwidth]{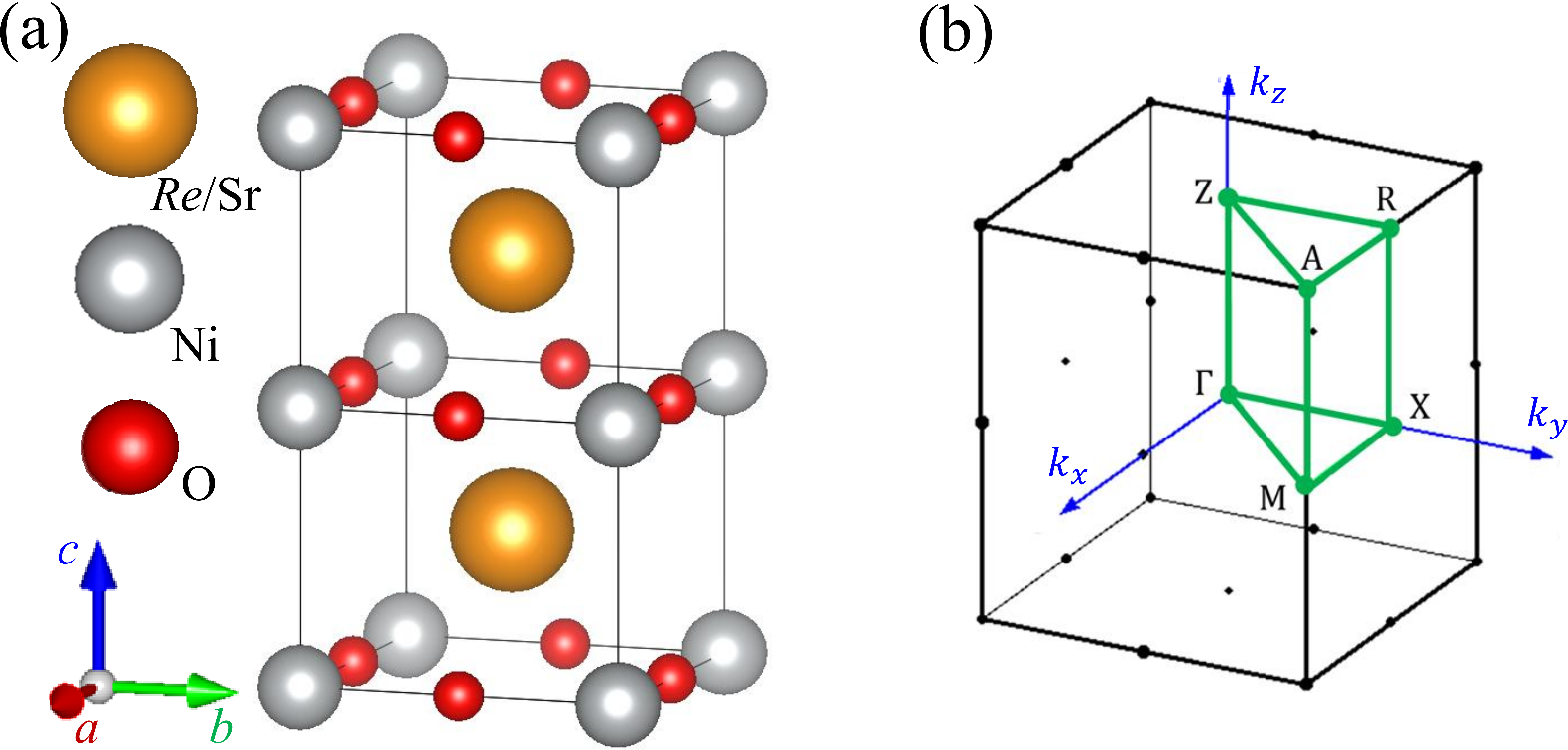}
\includegraphics[width=0.85\columnwidth]{NdSrNiO2Fig-band-DOS.eps}
\includegraphics[width=0.92\columnwidth]{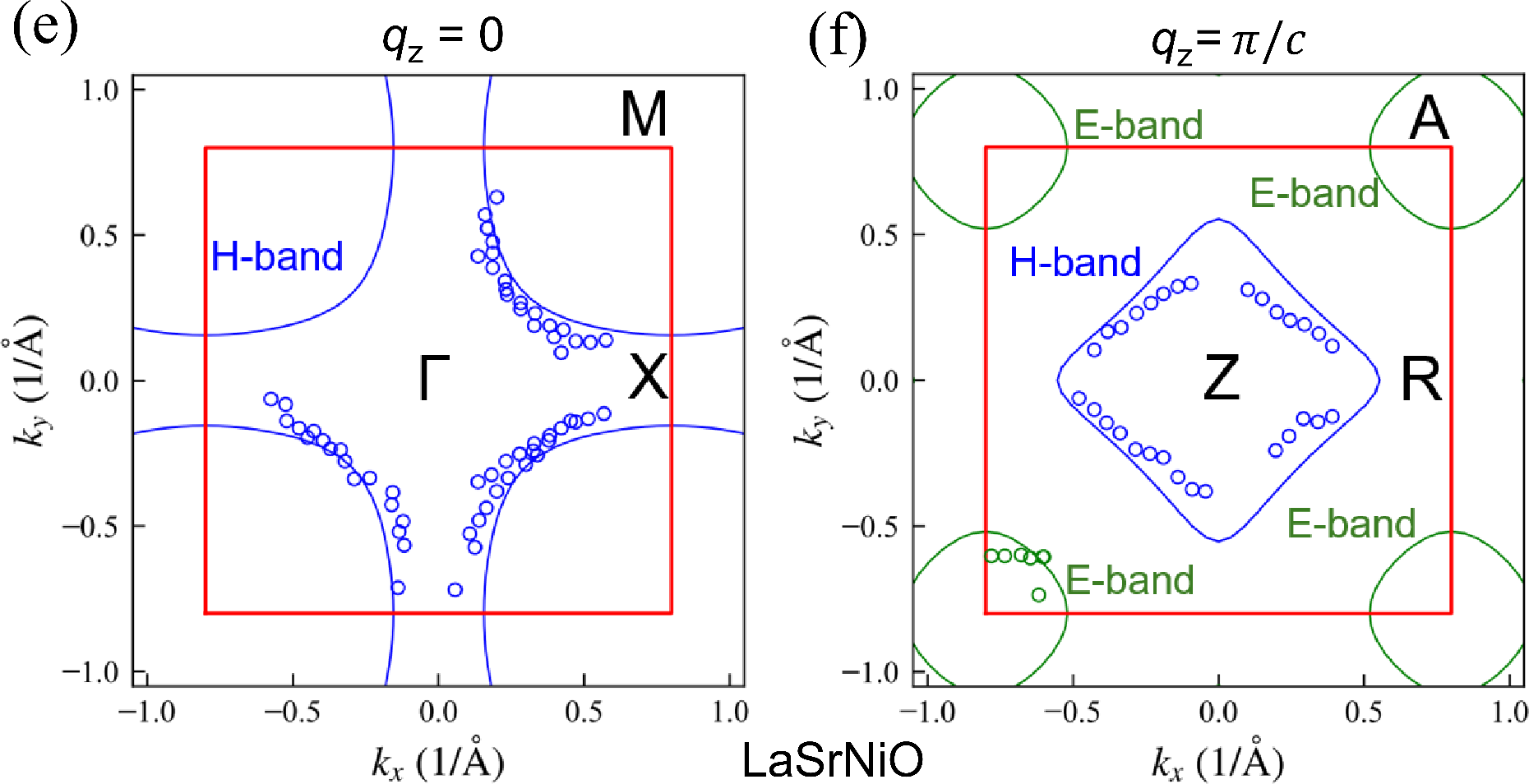}
\caption{Crystal and electronic structures. 
(a,b) Crystal structure and  Brillouin zone of the infinite-layer nickelates 
$Re$SrNiO ($Re =$ La, Pr, Nd).
(c,d) Ni $d_{x^2-y^2}$, Nd $d_{z^2}$ and O $p$ orbital-projected energy bands,
together with total and atom-decomposed density of states (DOS) of NdSrNiO.
In (c), atomic orbital weights are proportional to the circle sizes.
(e,f) Fermi surface (FS) in the $k_x-k_y$ plane at (e) $k_z = 0$ 
and (f) $k_z = \pi/c$ of LaSrNiO.
Solid curves represent the calculated FS, and open circles 
denote the ARPES measurements~\cite{Sun2025}.}
\label{fig:crystal}
\end{figure}

$Re_{1-x}$Sr$_x$NiO$_2$ ($Re =$ La, Pr, Nd) crystallizes in the layered tetragonal structure with
space group $P4/mmm$ (No. 123) and point group $D_{4h}$.
The crystal structure (Fig.~\ref{fig:crystal})
consists of alternating NiO$_2$ and $Re_{1-x}$Sr$_{x}$ layers.~\cite{Li2019,Hayward2003}
Its unit cell contains one formula unit (f.u.).~\cite{Li2019,Hayward2003}
The Wyckoff positions of Ni, Nd$_{1-x}$Sr$_x$, and O atoms are 1$a$ (0, 0, 0), 1$d$ (1/2, 1/2, 1/2),
and 2$f$ (1/2, 0, 0), respectively.
The central component is the NiO$_2$ layer, where Ni atoms form a two-dimensional (2D) square lattice
with O atoms sitting at the center of each bond between two neighboring Ni atoms.
Each Ni atom is thus coordinated by four in-plane O atoms.

The electronic structure calculations are based on the DFT
with the local density approximation~\cite{Perdew1981}.
The phonon dispersion and EPC matrix elements are calculated using the DFPT.~\cite{Baroni2001}
The plane wave pseudopotential method is used. The ultrasoft pseudopotentials~\cite{Corso2014}
are taken from the PSlibrary~\cite{Pslibrary}.
All these first-principles calculations are carried out using the QUANTUM ESPRESSO
package.~\cite{Giannozzi2009,Giannozzi2017}
Throughout this work, the highly efficient optimized tetrahedron
method for Brillouin zone integration~\cite{Kawamura2014} is adopted.
We use the SCTK code~\cite{Kawamura2020,sctk} to solve the SCDFT gap equation [Eq. (1)]
and to calculate the SC properties of $Re$SrNiO.
The calculated physical quantities on the Fermi surface (FS) are displayed
using the FermiSurfer program~\cite{Kawamura2019}.

\begin{figure}
\centering
\includegraphics[width=80mm]{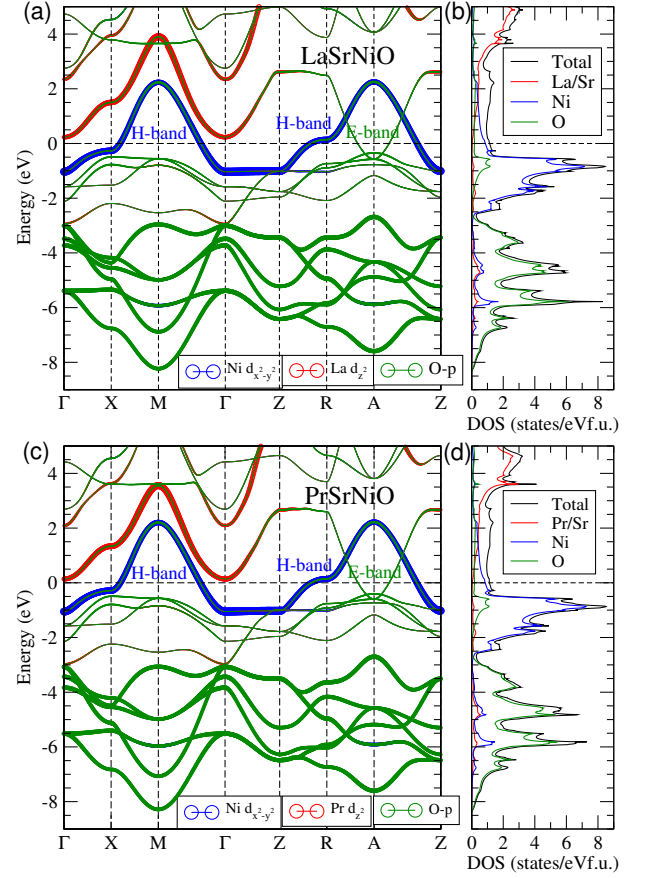}
\caption{Electronic structures of LaSrNiO and PrSrNiO.
(a,b) Ni $d_{x^2-y^2}$, La $d_{z^2}$ and O $p$ orbital-projected energy bands, total and atom-decomposed
density of states (DOS) of (a,b) LaSrNiO and (c,d) PrSrNiO.
In (a,c), atomic orbital weights are proportional to the circle sizes.}
\label{fig:LaPr_bands}
\end{figure}

\section{Electronic Structure and Fermi Surface}
\label{sec:Electronic_Structure}
\begin{figure}[!htb]
\centering
\includegraphics[width=0.85\columnwidth]{ReSrNiO2Fig-gapvsT.eps}
\includegraphics[width=0.92\columnwidth]{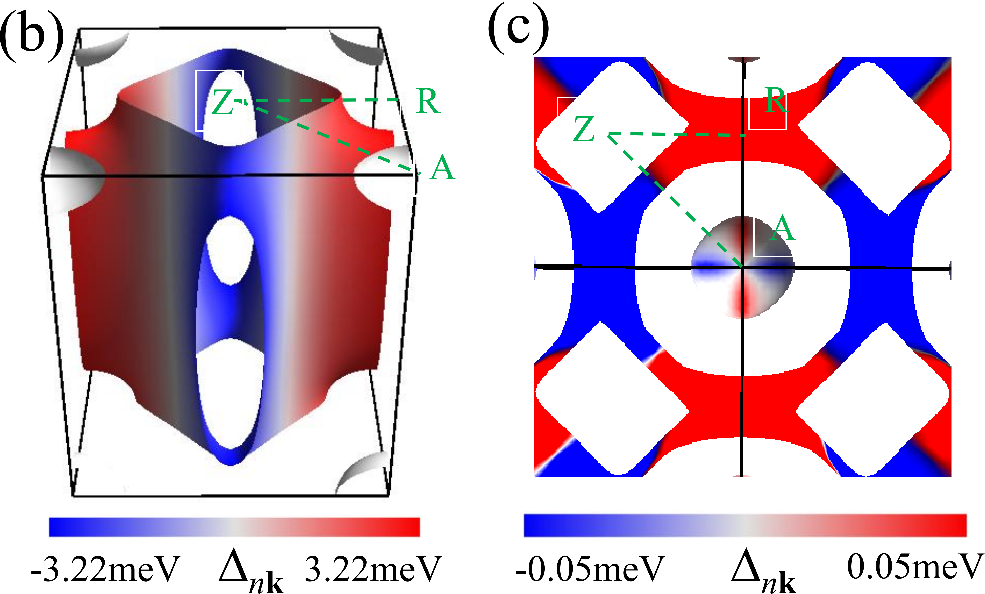}
\caption{Superconducting gap functions.
(a) Averaged (solid symbols) and maximum (open symbols) superconducting (SC) gap $\Delta$ of $Re$SrNiO
as a function of temperature ($T$). {\bf b},{\bf c}, band $n$ and momentum ${\bf k}$-dependent SC
gap $\Delta_{n{\bf k}}$ on the Fermi surface of NdSrNiO at 0.1 K in 
(b) the first Brillouin zone (BZ) and (c) the extended BZ and top view. 
The averaged $\Delta$ values are fitted with a BCS-type $T$-dependence.~\cite{Bardeen1957}
The fitting curves are plotted as solid lines, which are nearly identical to the calculated $\Delta$ values.
In (c), a smaller gap value scale is used so that
the $d_{x^2-y^2}$-wave gap function on the small FS pocket at the A point can be clearly seen.
}
\label{fig:gap_FS}
\end{figure}

The calculated electronic energy bands and density of states (DOS) of NdSrNiO are displayed in 
Fig.~\ref{fig:crystal}, and that of LaSrNiO and PrSrNiO are shown in Fig. \ref{fig:LaPr_bands}.
Since the electronic structure and physical properties of all three nickelates are similar 
(see Figs.~\ref{fig:crystal} and \ref{fig:LaPr_bands}), 
for simplicity, we focus on NdSrNiO unless otherwise stated. 
Figure ~\ref{fig:crystal}(c) shows that two bands (labelled H and E) cross the Fermi level ($E_F$). 
Consequently, the Fermi surface (FS) consists of a large quasi-two-dimensional cylinder
at the Brillouin zone (BZ) center and eight small three-dimensional electron pockets
sitting at the BZ corners (A points) (see Fig.~\ref{fig:crystal} and Fig.~\ref{fig:gap_FS}). 
Interestingly, in the $k_z = 0$ plane, the large quasi-two-dimensional FS sheet forms four 
circular hole pockets at the M points [Fig.~\ref{fig:crystal}(e)], similar to those 
in single-layered cuprate HgBa$_2$CuO$_{4+\delta}$~\cite{Sakakibara2012}. 
In contrast, in the $k_z = \pi/c$ plane, it transforms into a square electron pocket at the Z point 
(Fig.~\ref{fig:crystal}(f) and Fig. ~\ref{fig:gap_FS}), similar to those in 
single-layered cuprate La$_{2-x}$(Sr/Ba)$_x$CuO$_4$~\cite{Sakakibara2012}. 

Figure~\ref{fig:crystal}(d) indicates that in the region from -1.2 eV to 2.3 eV,
the DOS of $Re$SrNiO ($Re = $ La, Pr, Nd) comes primarily from the contributions of the Ni atoms.
Figure ~\ref{fig:crystal}(c) shows that the H band arises mainly from Ni 3$d_{x^2-y^2}$ orbital 
while the E band is dominated by O 2$p$ orbital with contributions from Nd 5$d$ orbitals. 
Calculated total and atom-decomposed DOS at the Fermi level ($N_F$) of $Re$SrNiO 
are listed in Table I, which indicates that the $N_F$ is dominated by contributions
from the Ni atoms with 15 \% $\sim$ 19 \% from the O atoms and only 3 \% from $Re$/Sr atoms.
Also listed in Table I are the bare Pauli spin susceptibility $\chi_0$ and the bare linear specific heat (Sommerfeld)
coefficient $\gamma_n$, which are related to the total DOS $N_F$ by $\chi_0 = (\mu_B)^2 N_F$
and $\gamma_n =  \frac{\pi^2}{3}k^{2}_{B}N_F$, respectively.
Our calculated energy bands and DOS spectra (Fig.~\ref{fig:crystal}) are 
similar to previously reported DFT calculations (see, e.g., ~\cite{Nomura2019,Li2024}). 
Importantly, our FS on the $k_z = 0$ and $k_z = \pi/c$ planes for LaSrNiO
agree well with the angle-resolved photoemission spectroscopy (ARPES) experiments~\cite{Sun2025}
[see Figs.~\ref{fig:crystal}(e) and \ref{fig:crystal}(f)]. 
No ARPES experiment on other $Re_{0.8}$Sr$_{0.2}$NiO$_2$ has been reported.\\ 

\begin{table}[b]
\caption{\label{tab:electronic}%
Calculated total and atom-decomposed density of states (DOS) at the Fermi level ($N_F$) of
infinite-layer nickelates $Re$SrNiO.  Also listed here are bare Pauli spin susceptibility 
$\chi_0$ and bare linear specific heat (Sommerfeld)
coefficient $\gamma_n$, which are related to the total DOS $N_F$ by $\chi_0 = (\mu_B)^2 N_F$
and $\gamma_n =  \frac{\pi^2}{3}k^{2}_{B}N_F$ , respectively.
}
\begin{ruledtabular}
\begin{tabular}{cccc}
                            & LaSrNiO         & PrSrNiO         &  NdSrNiO  \\
\hline
$N_F$ (states/eV/f.u.)      & 1.198           & 1.187           & 1.181 \\
Re/Sr (states/eV/f.u.)      & 0.037           & 0.036           & 0.034 \\
Ni    (states/eV/f.u.)     & 0.859           & 0.864           & 0.84  \\
O     (states/eV/f.u.)     & 0.176           & 0.177           & 0.222 \\
$\chi_0 $ ($10^{-5}$ emu/mol) & 4.01         & 3.98            & 3.96  \\
$\gamma_n$ (mJ/(mol-K$^2$))    & 2.82         & 2.80            & 2.78 \\
\end{tabular}
\end{ruledtabular}
\end{table}

\section{Superconducting properties}
\label{sec:SC}
We solve the SC gap equation [Eq.~(\ref{eq:gap-eq})] for many temperatures
in order to determine the SC transition temperature $T_c$.
In Fig.~\ref{fig:gap_FS}(a), the averaged and maximum SC gap values 
of $Re$SrNiO are plotted as a function of temperature ($T$).  
We fit the averaged $\Delta$ values with a BCS-type 
$T$-dependence $\Delta (T) = \Delta(0)\mathrm{tanh(\alpha\sqrt{{\it T_c}/{\it T}-1})}$~\cite{Bardeen1957}
by varying parameters $\Delta(0)$, $\alpha$ and $T_c$.
The obtained $\Delta(0)$ and $T_c$ values for all three nickelates are listed in Table II, and the
obtained $\alpha$ values are 1.67, 1.66 and 1.66 for LaSrNiO, PrSrNiO and NdSrNiO, respectively.
Importantly, Table II shows that our predicted $T_c$ values for all three $Re$SrNiO
agree well with the experimental values (see ~\cite{Chow2025} and references therein).

\begin{figure}[!htb]
\centering
\includegraphics[width=80mm]{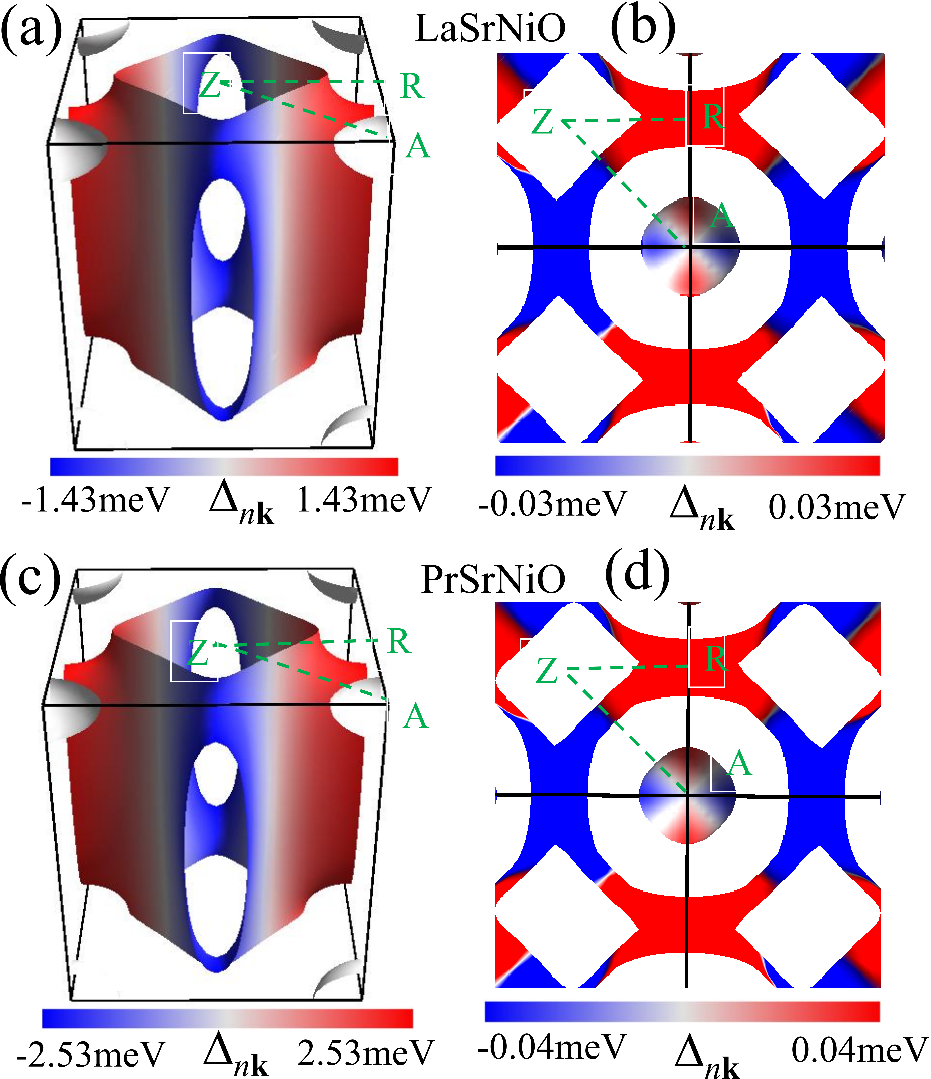}
\caption{Superconducting gap functions of LaSrNiO and PrSrNiO. 
Band $n$ and momentum ${\bf k}$-dependent superconducting (SC) gap $\Delta_{n{\bf k}}$
on the Fermi surface (FS) of (a,b) LaSrNiO and (c,d) PrSrNiO, in
(a,c) the first Brillouin zone and (b,d) the extended Brillouin zone.
Note that in (b,d), a smaller SC gap value scale is used so that the $d_{x^2-y^2}$-wave gap
function on the small FS pockets at the A points can show up clearly. }
\label{fig:LaPr-gap-FS}
\end{figure}

\begin{figure}[!htb]
\centering
\includegraphics[width=80mm]{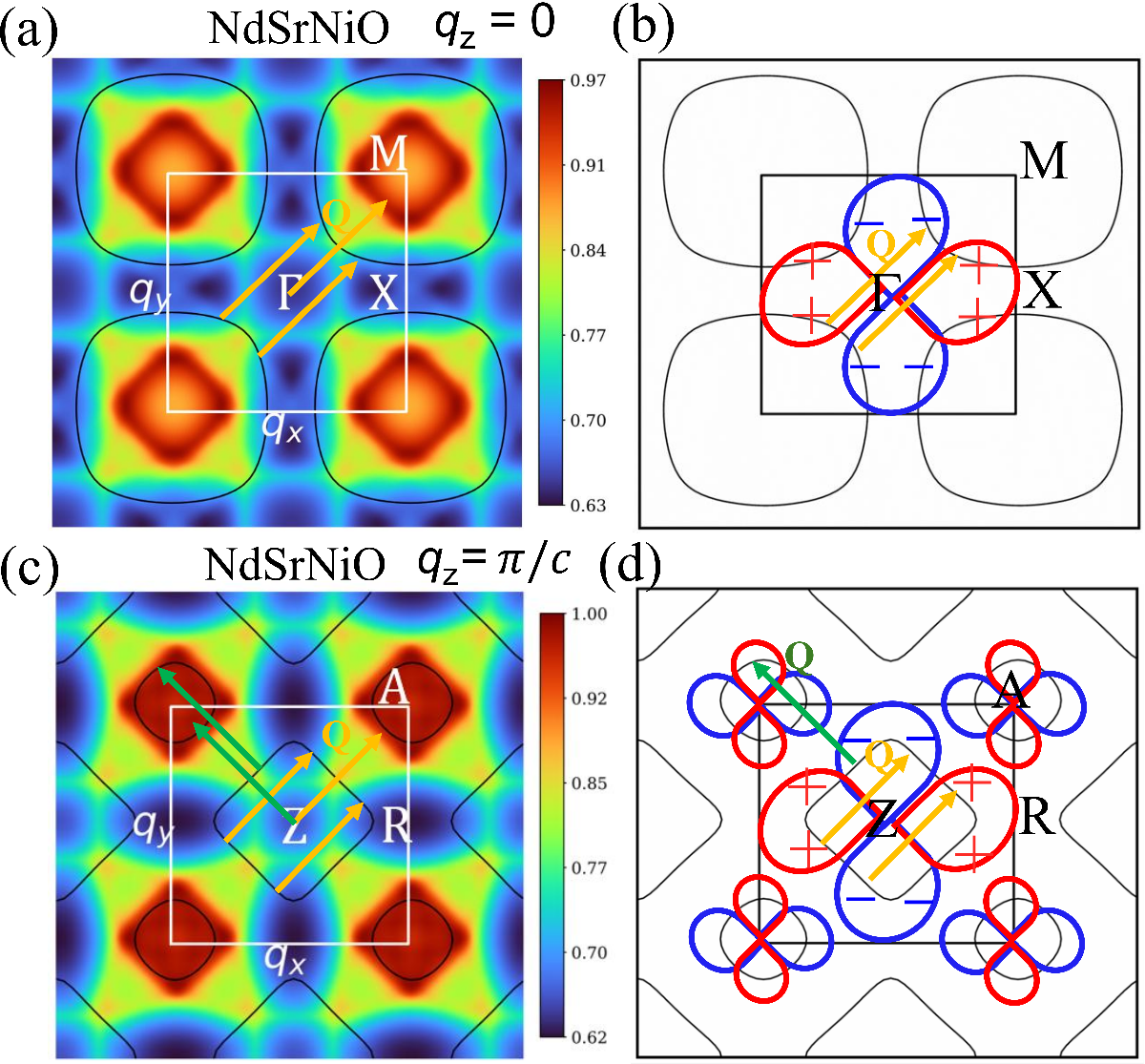}
\caption{Lindhard response function. Calculated bare susceptibility $\chi^0 ({\bf q})$
of NdSrNiO on the $q_x-q_y$ plane at (a) $q_z = 0$ and (c) $q_z = \pi/c$.
The displayed $\chi^0 ({\bf q})$ is normalized to its maximum value of 12.4.
The normalized minimum value of $\chi^0 ({\bf q})$ is 0.62.
Black curves represent the cross-sections of the Fermi surface (FS) pockets.
In (a) and (c), the three orange arrows represent equivalent nesting vectors Q along the [1,1] direction, 
at which $\chi^0 ({\bf q})$ reaches its maximum. These wave vectors connect FS regions with opposite signs 
of the superconducting gap, i.e., $\Delta_{\bf k} = -\Delta_{{\bf k} + {\bf Q}}$
(see (b) and (d) as well as Sec. VI). Similarly, there are two identical green arrows
pointing along the [-1,1] direction in (c), one indicating the wavevector ${\bf Q}$ at which $\chi^0 ({\bf q})$
peaks and the other denoting the inter-pocket SF pairing interaction, which results in sign change
inter-pocket $d_{x^2-y^2}(\pm)$-waves [see (d)].}
\label{fig:Lindhard}
\end{figure}

We now examine the symmetry and structure of the obtained SC gap function.
Band $n$ and momentum ${\bf k}$-dependent gap function $\Delta_{n{\bf k}}$
on the FS at $ T = 0.1$ K is displayed for NdSrNiO in Fig.~\ref{fig:gap_FS}
and for LaSrNiO and PrSrNiO in Fig. \ref{fig:LaPr-gap-FS}.
Remarkably, Fig.~\ref{fig:gap_FS} and Fig. \ref{fig:LaPr-gap-FS}
show that in all three nickelates $Re$SrNiO, the gap functions $\Delta_{n{\bf k}}$ 
on the large H-band FS sheet are $d_{x^2-y^2}$-waves with rather large magnitudes,
thus belonging to the spin-singlet even-parity pairing states with the $B_{1g}$ 
symmetry (see Table III in Appendix A  and ~\cite{Annett1990,Yip1993}). 
The gap functions $\Delta_{n{\bf k}}$ on the small 
E-band FS pockets are also $d_{x^2-y^2}$-waves, albeit with much smaller gap sizes 
[see Fig. ~\ref{fig:gap_FS}(c), Figs. \ref{fig:LaPr-gap-FS}(b) and \ref{fig:LaPr-gap-FS}(d)]. 
Moreover, the $d_{x^2-y^2}$-waves on the E-band FS pockets exhibit an opposite sign relative 
to those on the H-band FS sheets. Thus, our {\it ab initio} calculations
reveal that all three nickelates are two-band superconductors 
characterized by sign-changing $d_{x^2-y^2}$ gaps on disconnected FS pockets
[i.e., $d_{x^2-y^2}(\pm)$ gap structure with $B_{1g}$ symmetry].

For a weak-coupling isotropic $s$-wave phonon-mediated superconductor,
the SC gap-to-critical temperature ratio $\Delta_{ave}(0)/k_BT_c$ is 1.76 (the BCS value)~\cite{Bardeen1957}.
For strong-coupling $s$-wave phonon-mediated superconductors, 
$\Delta_{ave}(0)/k_BT_c = 1.76[1+12.5(T_c/\omega_{ln})^2ln(\omega_{ln}/2T_c)] \geq 1.76$ ~\cite{Carbotte1990},
where $\omega_{ln}$ is the logarithmically average phonon frequency (see Appendix B and Table II).
Table II shows that the calculated $\Delta_{ave}(0)/k_BT_c$ of $\sim$1.3
for all three nickelates is significantly lower than the BCS value~\cite{Bardeen1957},
further supporting the unconventional nodal superconductivity revealed by the momentum-dependent 
gap functions in Figs. 3 and 4.
For an unconventional nodal superconductor,
the gap-to-$T_c$ ratio $\Delta_{max}(0)/k_BT_c$ is expected to be significantly greater 
than the BCS value~\cite{Sigrist2005}. Indeed, Table II indicates that 
the calculated $\Delta_{max}(0)/k_BT_c$ for all three nickelates is $\sim$2.3, 
substantially exceeding the BCS value and thus providing further evidence for nodal superconductivity.

\begin{figure}
\centering
\includegraphics[width=80mm]{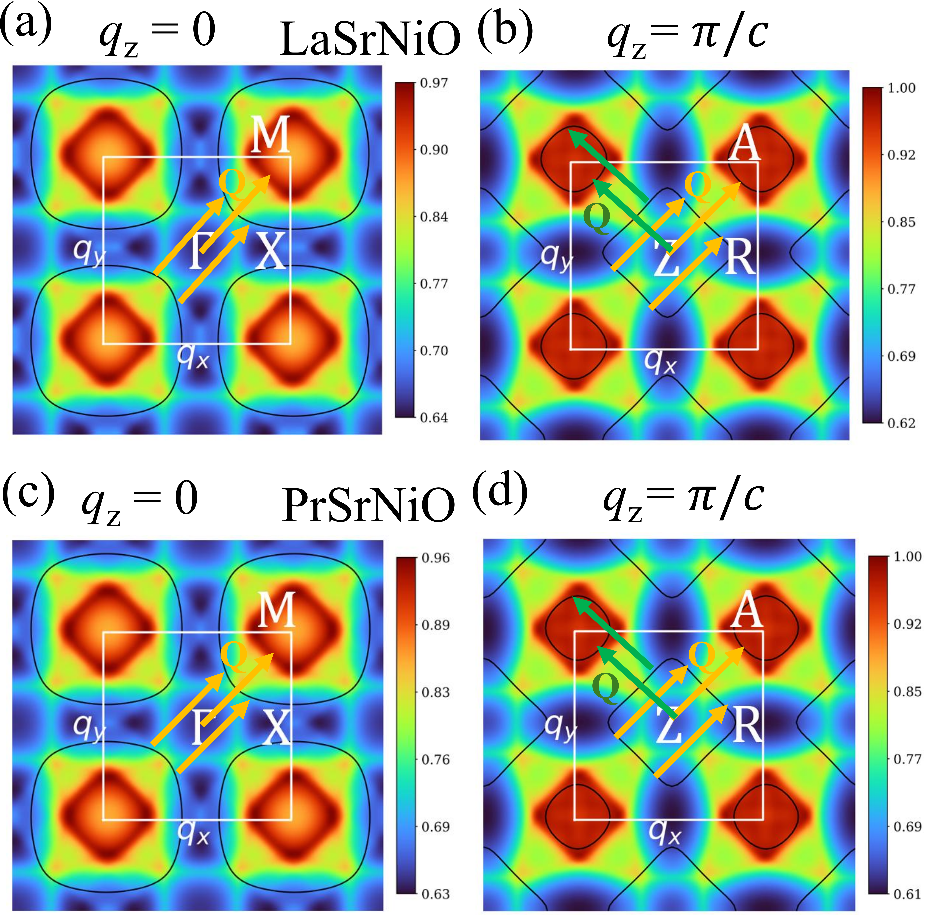}
\caption{Lindhard response function.
Calculated bare susceptibility $\chi^0 ({\bf q})$ of
LaSrNiO (a,b) and PrSrNiO (c,d) on the $q_x-q_y$ plane
at (a,c) $q_z = 0$ and (b,d) $q_z = \pi/c$. The displayed $\chi^0 ({\bf q})$ is normalized to 
its maximum value of 12.55 for LaSrNiO and of 12.54 for PrSrNiO.
The normalized minimum value of $\chi^0 ({\bf q})$ is 0.62 for LaSrNiO and is 0.61 for PrSrNiO.
Black curves represent the cross-sections of the hole and electron Fermi surface (FS) pockets.
In each panel, there are three orange arrows represent equivalent nesting vector ${\bf Q}$ 
along the [1,1] direction, at which $\chi^0 ({\bf q})$ reaches its maximum.
These wave vectors connect FS regions with opposite signs of the superconducting gaps, i.e., 
$\Delta_{\bf k} = -\Delta_{\bf k + Q}$ [see Figs. ~\ref{fig:Lindhard}(b) and ~\ref{fig:Lindhard}(b) as well as Sec. VI]. 
Similarly, there are two identical green arrows pointing
along the [-1,1] direction in (b) and (d), one indicating the wavevector ${\bf Q}$
at which $\chi^0 ({\bf q})$ peaks and the other denoting the inter-pocket SF pairing interaction,
which results in sign-change inter-pocket $d_{x^2-y^2}(\pm)$-waves [see Fig. ~\ref{fig:Lindhard}(b)].}
\label{fig:Lindhard-LaPr}
\end{figure}

\section{Origin and mechanism of the unconventional Superconductivity}
\label{sec:origin}
To understand the origin of the unconventional $d$-wave superconductivity identified in the preceding section,
we perform the following "computer experiments".
First, we turn off both the SF and screened Coulomb repulsion
(i.e., purely phonon-mediated pairing mechanism; see Appendix B for details). 
The calculated $T_c$ reduces to $\sim$1.0 K 
for all the nickelates (see Table II) and $Re$SrNiO 
become a conventional phonon-mediated $s$-wave superconductor with the gap $\Delta_{n{\bf k}}$
on the FS varying from 0.07 to 0.25 meV (see Fig. 10 in Appendix B). 
This is similar to the case of $\gamma$-BiPd, which is a single gap anisotropic $s$-wave superconductor~\cite{Keshri2025}.
The $s$-wave superconductivity in this case is further supported by the fact that
the calculated gap-to-critical temperature ratio $\Delta_{ave}(0)/k_BT_c$ of $\sim$1.7 
for all three nickelates (see Table II) is close to the BCS value of 1.76 for the weak-coupling
phonon-mediated $s$-wave superconductors~\cite{Bardeen1957}.
Second, if we turn on the screened Coulomb repulsion, the weak phonon-mediated superconductivity 
is suppressed with $T_c$ of $\sim$0.01 K (Table II).
These "computer experiments" thus demonstrate clearly that the superconductivity in $Re$SrNiO
is unconventional and arises from the SF-mediated pairing mechanism~\cite{Scalapino1999}.

\begin{table*}
\caption{
\label{tab:SC-property}%
Calculated superconducting (SC) properties of infinite-layer nickelates $Re$SrNiO for
(a) electron-phonon coupling (EPC) only, (b) EPC plus screened Coulomb repulsion ($\mu$) (EPC + $\mu$),
and (c) EPC plus $\mu$ and spin fluctuation (SF) (EPC+$\mu$+SF), namely, transition temperature ($T_c$), 
averaged ($\Delta_{ave}$) and maximum ($\Delta_{max}$) SC gap, logarithmically averaged phonon frequency ($\omega_{ln}$),
EPC constant ($\lambda$), screened Coulomb repulsion ($\mu$) and renormalization ($Z$). 
The experimental transition temperatures ($T_c^{exp}$) are taken from ~\cite{Chow2025}
and placed in the bracket in the second column.}
\begin{ruledtabular}
\begin{tabular}{cccccccccc}
                 &                   &\multicolumn{6}{c}{LaSrNiO}&       &       \\ \hline
    & $T_{c}$ ($T_{c}^{exp}$) (K) & $\Delta_{ave}$ (meV) & $\Delta_{max}$ (meV) & $\Delta_{ave}/k_BT_{c}$ & $\Delta_{max}/k_BT_{c}$ & $\omega_{ln}$ (K) & $\lambda$ & $\mu$  & $Z$ \\
(a) EPC          & 0.91       & 0.135  & 0.213   &  1.72  &  2.72  & 371 & 0.143     & 0.0    & 0.154   \\
(b) EPC+$\mu$    & 0.0        & 0.003  & 0.006   &  0.0   &  0.0   & 371 & 0.143     & 0.390  & 0.953   \\
(c) EPC+$\mu$+SF & 7.3 (9)    & 0.808  & 1.427   &  1.28  &  2.27  & 371 & 0.143     & 2.968  & 2.276   \\
\hline
                 &                   &\multicolumn{6}{c}{PrSrNiO}&       &        \\ \hline
    & $T_{c}$ ($T_{c}^{exp}$) (K) & $\Delta_{ave}$ (meV) & $\Delta_{max}$ (meV) & $\Delta_{ave}/k_BT_{c}$ & $\Delta_{max}/k_BT_{c}$ & $\omega_{ln}$ (K) & $\lambda$ & $\mu$  & $Z$ \\
(a) EPC          & 1.077      & 0.158 & 0.254    &  1.70  &  2.74  & 346 & 0.153     & 0.0    & 0.160   \\
(b) EPC+$\mu$    & 0.0        & 0.004 & 0.0      &  0.0   &  0.0   & 346 & 0.153     & 0.388  & 0.962   \\
(c) EPC+$\mu$+SF & 12.7 (12)  & 1.401 & 2.530    &  1.28  &  2.31  & 346 & 0.153     & 3.053  & 2.336   \\
\hline
                 &                   &\multicolumn{6}{c}{NdSrNiO}&       &       \\ \hline
    & $T_{c}$ ($T_{c}^{exp}$) (K) & $\Delta_{ave}$ (meV) & $\Delta_{max}$ (meV) & $\Delta_{ave}/k_BT_{c}$ & $\Delta_{max}/k_BT_{c}$ & $\omega_{ln}$ (K) & $\lambda$ & $\mu$  & $Z$ \\
(a) EPC          & 1.25       & 0.178 & 0.280    &  1.65  &  2.63  & 303 & 0.142     & 0.0    & 0.145   \\
(b) EPC+$\mu$    & 0.01       & 0.0   & 0.0      &  0.0   &  0.0   & 303 & 0.142     & 0.387  & 0.944   \\
(c) EPC+$\mu$+SF & 16.1 (15)  & 1.768 & 3.247    &  1.18  &  2.34  & 303 & 0.142     & 3.100  & 2.331   \\
\end{tabular}
\end{ruledtabular}
\end{table*}

Detailed insight into the SC pairing mechanism can be gained by examining band $n$- and momentum ${\bf k}$-dependent
EPC strength ($\lambda_{n\bf{k}}$), screened Coulomb repulsion ($\mu_{n\bf{k}}^{ee}$)
and SF-induced  pairing interaction ($\mu_{n\bf{k}}^{SF}$) on the FS of the nickelates,
which are presented in Fig. ~\ref{fig:lambda_FS} in Appendix C.
First, Fig. ~\ref{fig:lambda_FS} indicates that $\mu_{n\bf{k}}^{ee}$
is always larger than  $\lambda_{n\bf{k}}$ on the large H-band FS sheet, while the opposite
is true on the small E-band FS pockets. Because the signs of $\mu_{n\bf{k}}^{ee}$ and $\lambda_{n\bf{k}}$ 
are opposite, the screened Coulomb repulsion suppresses the phonon-mediated attraction
on the H-band FS sheet, while the phonon-mediated attraction dominates on the E-band FS pocket.
Since the surface area of the H-band sheet is much larger than that of the E-band pocket,
this explains why the weak phonon-induced superconductivity in $Re$SrNiO
is suppressed when the Coulomb repulsion is switched on.
Second,  Fig.~\ref{fig:lambda_FS} shows that $\mu_{n\bf{k}}^{SF}$ is ten times stronger than 
both $\mu_{n\bf{k}}^{ee}$ and  $\lambda_{n\bf{k}}$ on the large H-band FS sheet. 
This results in the dominating SF-mediated e-e pairing and SF-driven unconventional 
superconductivity in $Re$SrNiO.

To provide an intuitive picture of the $d_{x^2-y^2}$-wave superconductivity and 
SF-mediated pairing mechanism identified in this work,
we depict the calculated bare susceptibility (Lindhard response function) 
\begin{align}
\chi^0({\bf q}) = -\sum_{nm{\bf k}}\frac{\theta(\varepsilon_F-\varepsilon_{m{\bf k}+{\bf q}})-\theta(\varepsilon_F-\varepsilon_{n{\bf k}})}{\varepsilon_{m{\bf k}+{\bf q}}-\varepsilon_{n{\bf k}} }
\end{align}
in Fig.~\ref{fig:Lindhard} and Fig. \ref{fig:Lindhard-LaPr}.
These figures show that in the $q_z = 0$ plane, there is a pronounced square-shape susceptibility peak
near the corner M point due to the inter-FS pocket nesting as indicated by the orange arrows.
The effective interaction between electronic states ${\bf k}$ and ${\bf k}' = {\bf k}+{\bf q}$ mediated 
by spin fluctuations is given by ~\cite{Sigrist2005}
\begin{align}
V_{eff}({\bf k},{\bf k}') = \frac{3I^2}{4}\chi_{SF}({\bf q})=\frac{3I^2}{4}\frac{\chi^{0}({\bf q})}{1-I\chi^{0}({\bf q})}
\label{effective interaction}
\end{align}
where $I$ is the Stoner exchange interaction parameter~\cite{Janak1977}.
Therefore, a peak near the wavevector ${\bf q} \approx {\bf Q}$ in susceptibility $\chi^{0}({\bf q})$ would lead 
to a strong antiferromagnetic (AF) interaction $V_{eff}({\bf k},{\bf k}')$ between two electrons having wavevectors ${\bf k}$
and ${\bf k}' = {\bf k}+{\bf Q}$. Moreover $V_{eff}({\bf k},{\bf k}')$ is repulsive.
Consequently, an isotropic $s$-wave pairing state cannot satisfy  
the BCS gap equation ($T = 0$ K)~\cite{Scalapino1999}
\begin{align}
\Delta_{\bf k} = - \sum_{{\bf k}'}V_{eff}({\bf k},{\bf k}')\frac{\Delta_{{\bf k}'}}{2E_{{\bf k}'}}.
\end{align}
However, if the gap function changes sign, $\Delta_{\bf k} = - \Delta_{{\bf k} + {\bf Q}}$,
a solution would be possible. In the present case, $\chi^0({\bf q})$ peaks near ${\bf Q} = (\pi/a, \pi/a)$.
As a result, due to inter-FS pocket AF coupling indicated by the orange arrows in Fig. ~\ref{fig:Lindhard}(a), 
Fig. ~\ref{fig:Lindhard}(c), and Fig. \ref{fig:Lindhard-LaPr}, $d_{x^2-y^2}$-wave pairing 
would occur, as illustrated in Figs. ~\ref{fig:Lindhard}(b) and ~\ref{fig:Lindhard}(d).
Similarly, in the $q_z = \pi/c$ plane, there is a prominent square plateau
at A due to the intra-FS pocket nesting, as indicated by the orange arrows
in Fig. ~\ref{fig:Lindhard}(b), Figs. \ref{fig:Lindhard-LaPr}(b) and \ref{fig:Lindhard-LaPr}(d).
This also gives rise to the $d_{x^2-y^2}$-wave pairing. 
Furthermore, these $\chi^0({\bf q})$ peaks near ${\bf Q} = (\pi/a, \pi/a)$
would also induce inter-FS pocket AF coupling between the H-band FS pocket at Z
and the E-band FS pocket at A [see the green arrows in Fig. ~\ref{fig:Lindhard}(c), 
Figs. \ref{fig:Lindhard-LaPr}(b) and \ref{fig:Lindhard-LaPr}(d)].
This would result in the inter-FS pocket $d_{x^2-y^2}$-wave pairing with signs reversed
[see Fig. ~\ref{fig:Lindhard}(d)].\\

\begin{figure}[!htb]
\centering
\includegraphics[width=0.85\columnwidth]{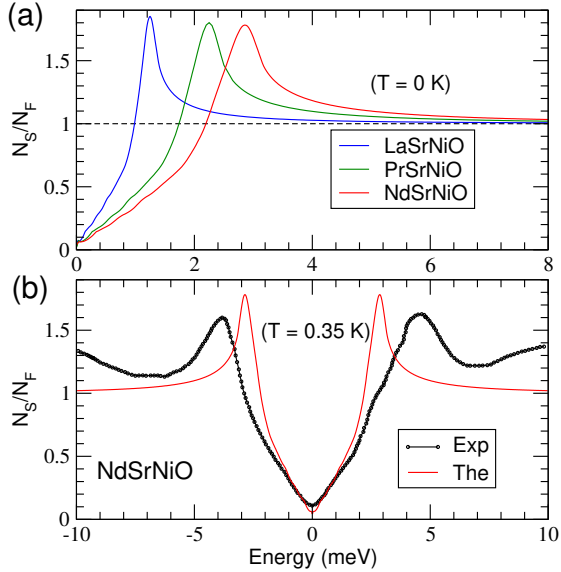}
\caption{Superconducting quasiparticle density of states.
(a) Normalized superconducting quasiparticle density of states (DOS) $N_S/N_F$ 
of $Re$SrNiO calculated as a function of excitation energy
at $T = 0$ K, where $N_F$ is the normal state DOS at $E_F$.
(b) Theoretical $N_S/N_F$ spectrum at $T = 0$ K (red curve) compared with
a typical STS $N_S/N_F$ spectrum (black curve) at $T = 0.35$ K taken on
a flat surface region of NdSrNiO films~\cite{Gu2020}. 
}
\label{fig:qpdos}
\end{figure}


\section{Comparison with experiments and discussion}
\label{sec:discussion}
Using the obtained SC gap function $\Delta_{n\bf{k}}$, we compute experimentally measurable SC properties. 
For example, the SC quasiparticle density of states (QPDOS) 
($N_s$) spectra of $Re$SrNiO could be probed by STS experiments. 
Indeed, STS measurements were recently performed on Nd$_{1-x}$Sr$_{x}$NiO$_2$ 
$(x = 0.12-0.25)$ thin films~\cite{Gu2020}. Since the nodeless and nodal SC gap functions exhibit 
different features in the QPDOS spectra, the measured QPDOS spectra would allow
us to study the nature of the SC gap, in particular, the nodeless $s$-wave or nodal 
$d$-wave (see ~\cite{Gu2020} and references therein).
The QPDOS can be written as~\cite{Kawamura2017}
\begin{equation}
N_s(\varepsilon)= \sum_{n{\bf k}}\delta(\varepsilon - E_{n{\bf k}}) 
= \sum_{n{\bf k}}\int d\xi\delta(\xi-\xi_{n{\bf k}})\delta(\varepsilon - E_{n{\bf k}}(\xi)),
\label{eq:qpdos}
\end{equation}
and is calculated using the optimized tetrahedron method~\cite{Kawamura2014}.
The normalized SC QPDOS ($N_s/N_F$) spectra of $Re$SrNiO 
calculated at $ T = 0$ K are plotted in Fig.~\ref{fig:qpdos}(a). 
Here $N_F$ denotes the normal state DOS at $E_F$.
Importantly, Fig.~\ref{fig:qpdos}(a) shows that all three nickelates have a V-shaped SC gap function,
indicating nodal SC gap structures in these nickelates, as 
can be expected from our predicted $d_{x^2-y^2}$-wave gap function 
in $Re$SrNiO reported in Sec. ~\ref{sec:SC} (see Fig. ~\ref{fig:gap_FS} and Fig. ~\ref{fig:LaPr-gap-FS}).
A typical STS spectrum measured at 0.35 K from a flat surface region of 
NdSrNiO thin films~\cite{Gu2020} is reproduced in Fig.~\ref{fig:qpdos}(b).
Figure ~\ref{fig:qpdos}(b) shows that the calculated and measured SC QPDOS spectra 
are in good agreement, especially in the vicinity of the zero excitation energy.
Nevertheless, a fully gapped feature can also be seen on the spectra taken on
a rough surface spot, implying nodeless $s$-wave superconductivity in the NdSrNiO film as well.~\cite{Gu2020} 
These inconsistent spatially varying gap structures observed in the NdSrNiO
films require further STS experiments on high quality samples or 
other types of experiments with bulk-sensitive probes. 
To date, no STS experiment on LaSrNiO and PrSrNiO has been reported. 
We are confident that our interesting work will stimulate such STS experiments on La$_{0.8}$Sr$_{0.2}$NiO$_2$ 
and Pr$_{0.8}$Sr$_{0.2}$NiO$_2$ in the near future.

Measurements of the $T$-dependence of the superfluid density $\rho_s(T)$ also provide information
on the pairing symmetry. 
In the clean limit, as $T$ approaches zero, a nodeless $s$-wave gap would result in an exponential
saturation in $\rho_s$, while a nodal $d$-wave gap would lead to a linear $T$ dependence~\cite{Prozorov2006}.  
Two recent measurements on $\rho_s(T)$ in Sr-doped nickelates via $T$-dependent magnetic penetration depth (MPD) 
experiments~\cite{Harvey2025,Chow2022} were reported. Harvey {\it et al.}~\cite{Harvey2025} found that 
their measured $\rho_s(T)$ suggests nodal superconductivity in LaSrNiO and PrSrNiO, 
while it exhibits a complex behavior in NdSrNiO due to Nd 4$f$ magnetism, 
which precludes a direct measure of the nodal structure. In contrast, Chow {\it et al.} reported that 
their results indicate that the SC order parameter in NdSrNiO is beyond a single $d_{x^2-y^2}$-wave gap.~\cite{Chow2022} 
These ambiguous conclusions on the SC pairing symmetry in NdSrNiO could be caused by either Nd 4$f$ magnetism 
or the MPD experiments being a surface sensitive technique. Recently, a terahertz spectroscopy (a bulk-sensitive probe) 
study on $\rho_s(T)$ in Nd$_{0.85}$Sr$_{0.15}$NiO$_2$~\cite{Cheng2024}
indicated $d$-wave superconductivity in the films. 
Note that the terahertz spectroscopy is a bulk-sensitive technique.
Thus, one could tentatively conclude that there is strong experimental evidence for nodal $d$-wave
superconductivity in Sr-doped infinite-layer nickelates at least in LaSrNiO and PrSrNiO, 
although further experiments on NdSrNiO 
would be needed to understand the complex $\rho_s(T)$ behavior in Nd$_{1-x}$Sr$_{x}$NiO$_2$ observed in
the MPD experiments~\cite{Harvey2025,Chow2022}.   

As mentioned before, recent {\it ab initio} GW and GWPT calculations~\cite{Li2024}
predicted that Nd$_{0.8}$Sr$_{0.2}$NiO$_2$ is a phonon-mediated two-gap superconductor with the
calculated $T_c$ in agreement with experiments. 
This prediction differs profoundly  
from the SF-driven $d_{x^2-y^2}$-wave pairing found in the present work.
The {\it ab initio} GW approach has achieved considerable success in describing quasiparticle properties of
solids including the EPC (see ~\cite{Li2024} and references therein). However, 
the SF pairing interaction was omitted and the screened e-e Coulomb repulsion was treated as a semiempirical
parameter $\mu^*$~\cite{Li2024}. We believe that whether the present SCDFT methodology or
the combined GW and GWPT and Eliashberg theory approach~\cite{Li2024} is more appropriate
for infinite-layer nickelates should be best judged by experiments. 
We notice that at least two distinct differences
in the predictions from the present and previous work~\cite{Li2024} could be tested experimentally.
First, the calculated energy bands near $E_F$ and thus the FS topology are very different
and this can be verified by the ARPES measurements on NdSrNiO.
Although our calculated FS cross-sections in LaSrNiO agree well
with the ARPES experiments~\cite{Sun2025} (see Fig.~\ref{fig:crystal}c), the band structure
of NdSrNiO could be different. Consequently, it will be very helpful to perform the ARPES experiments
on NdSrNiO as well as PrSrNiO. Second, the profound difference in the predicted SC gap structure could be
also examined by several different  kinds of experiments such as STS and MPD measurements.
Although the MPD experiments indicated nodal $d$-wave superconductivity in LaSrNiO and 
PrSrNiO~\cite{Harvey2025}, no conclusion could be made on NdSrNiO due to the complex behavior caused 
by Nd 4$f$ magnetism~\cite{Harvey2025}. 
High-resolution ARPES measurements~\cite{Hashimoto2014} may decisively determine the SC gap structure in NdSrNiO
and other nickelate superconductors. \\
 

\section*{Acknowledgments}
We thank Chia Ling Chien, Mitsuaki Kawamura, Zhenglu Li and Jau-Wen Liu for helpful discussions. 
We acknowledge the support from the National Science and Technology Council (NSTC)
and National Center for Theoretical Sciences (NCTS), Taiwan. We also thank the National Center
for High-performance Computing (NCHC) in Taiwan for the computing time. \\


\appendix
\section{Irreducible representations and basis functions for point group $D_{4h}$}
\label{sec:appendix_A}
The symmetry and structure of the SC gap function (i.e., the SC order parameter) of a superconductor
are not only determined by the types of the microscopic pairing-interactions but also 
by the crystalline symmetry of the superconductor.~\cite{Annett1990,Yip1993} 
For example, a centrosymmetric superconductor
can admit either spin-singlet pairing or spin-triplet pairing, i.e., coexistence of 
spin-singlet and spin-triplet SC states is forbidden. On the other hand, in a noncentrosymmetric
metal, both spin-singlet and spin-triplet pairing states can occur simultaneously.
Furthermore, the structure of the SC gap function is dictated by the underlying point group symmetry
of the superconductor.~\cite{Annett1990,Yip1993} For example, for a cubic superconductor, $d_{xy}$-wave
SC state is not allowed.~\cite{Yip1993}
Infinite-layer nickelates $Re$SrNiO crystalize in a centrosymmetric tetragonal structure with
the $D_{4h}$ point group symmetry. Table III lists the irreducible representations (IRREPs) and symmetry-allowed
basis functions as well as character tables for $D_{4h}$~\cite{Annett1990,Yip1993}. Even-parity basis functions are for
spin-singlet pairing states and odd-parity basis functions are for spin-triplet pairing states~\cite{Annett1990,Yip1993}.

\begin{table*}
\caption{\label{tab:symmetry}
Irreducible representations (IRREPs), basis functions and character tables for tetragonal crystals with
the $D_{4h}$ point group symmetry.~\cite{Annett1990,Yip1993}. Even-parity basis functions are for
spin-singlet pairing states and odd-parity basis functions are for spin-triplet pairing
states.~\cite{Annett1990,Yip1993} }
\begin{ruledtabular}
\begin{tabular}{cccccccccccc}
IRREP  & Even basis functions           &$E$ &$C_2$ &2$C_4$ &2$C_2^{'}$ &2$C_2^{"}$ &$i$ &$iC_2$ &2$iC_4$ &2$iC_2^{'}$ &2$iC_2^{"}$ \\ \hline
$A_{1g}$ & $1, (x^2 + y^2), (3z^2-r^2)$ & 1  & 1    & 1     & 1       & 1       & 1  & 1     & 1      & 1        & 1        \\
$A_{2g}$ & $x y (x^2 - y^2)$            & 1  & 1    & 1     & -1      & -1      & 1  & 1     & 1      & -1       & -1        \\
$B_{1g}$ & $x^2 - y^2$                  & 1  & 1    & -1    & 1       & -1      & 1  & 1     & -1     & 1        & -1       \\
$B_{2g}$ & $xy$                         & 1  & 1    & -1    & -1      & 1       & 1  & 1     & -1     & -1       & 1       \\
$E_g   $ & $z(x, y); z(x^3, y^3)$       & 2  & -2   & 0     & 0       & 0       & 2  & -2    & 0      & 0        & 0        \\ \hline
IRREP  & Odd basis functions            &$E$ &$C_2$ &2$C_4$ &2$C_2^{'}$ &2$C_2^{"}$ &$i$ &$iC_2$ &2$iC_4$ &2$iC_2^{'}$ &2$iC_2^{"}$ \\ \hline
$A_{1u}$ & $xyz(x^2 - y^2)$             & 1  & 1    & 1     & 1       & 1       & -1 & -1    & -1     & -1       & -1   \\
$A_{2u}$ & $z$                          & 1  & 1    & 1     & -1      & -1      & -1 & -1    & -1     & 1        & 1    \\
$B_{1u}$ & $xyz$                        & 1  & 1    & -1    & 1       & -1       & -1 & -1    & 1      & -1       & 1    \\
$B_{2u}$ & $z(x^2 - y^2)$               & 1  & 1    & -1    & -1      & 1      & -1 & -1    & 1      & 1        & -1   \\
$E_u   $ & $(x, y)$                     & 2  & -2   & 0     & 0       & 0       & -2 &  2    & 0      & 0        & 0    \\
\end{tabular}
\end{ruledtabular}
\end{table*}

\section{Phonon dispersion, electron-phonon coupling and phonon-mediated superconductivity}
\label{sec:appenix_B}

\begin{figure}
\centering
\includegraphics[width=80mm]{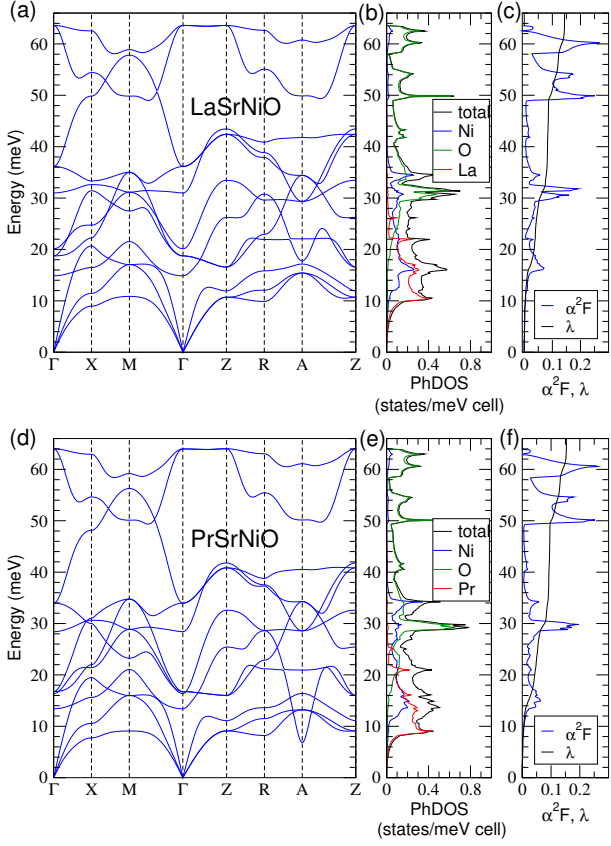}
\caption{(a,d) Phonon dispersion, (b,e) phonon density of states (PhDOS), (c,f) Eliashberg spectral
function ($\alpha^{2}F$) and accumulative electron-phonon coupling constant [$\lambda(\omega)$]
of (a,b,c) LaSrNiO and (d,e,f) PrSrNiO.}
\label{fig:LaPr-ph-bands}
\end{figure}

\begin{figure}
\centering
\includegraphics[width=80mm]{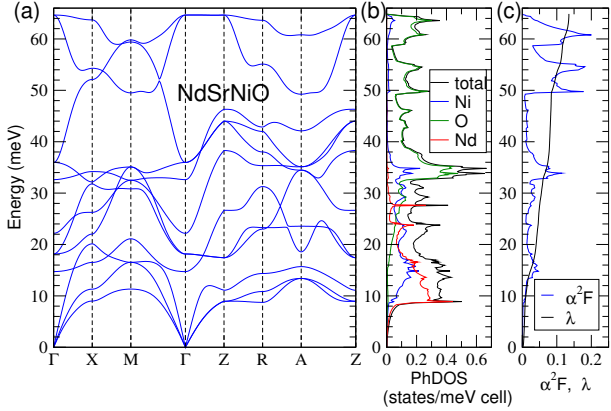}
\caption{(a) Phonon dispersion, (b) phonon density of states (PhDOS), (c) Eliashberg spectral
function ($\alpha^{2}F$) and accumulative electron-phonon coupling constant [$\lambda(\omega)$]
of NdSrNiO.}
\label{fig:Nd-ph-bands}
\end{figure}

Calculated phonon dispersion and phonon DOS (PhDOS) of LaSrNiO, PrSrNiO and NdSrNiO are presented
in Figs. ~\ref{fig:LaPr-ph-bands} and ~\ref{fig:Nd-ph-bands}.
Since their unit cell contains four atoms, $Re$SrNiO has 12 phonon modes (bands): 
three acoustic and nine optical ones, as shown in Figs. ~\ref{fig:LaPr-ph-bands} and ~\ref{fig:Nd-ph-bands}.
Below $\sim$20 meV, the PhDOS is dominated by the vibrations of heavy rare earth atoms (La, Pr and Nd).
On the other hand, light O atomic vibrations become dominant above $\sim$24 meV.
The vibrations of Ni atoms make significant contributions in the middle frequency range from 10 meV to 36 meV.
The $\Gamma$-point in the BZ exhibits centrosymmetric point group symmetry $D_{4h}(4/mmm)$.
Within this symmetry, the longitudinal acoustic phonon mode corresponds to the $A_{2u}$ representation,
while the two transverse acoustic modes belong to the $E_{1u}$ representation (see Table III).
In the acoustic region, the transverse modes exhibit higher frequencies than the longitudinal mode.
In contrast, in the optical region, the longitudinal optical modes have higher frequencies
than the transverse optical modes.

To better understand the phonon-mediated superconductivity, one usually calculates 
the isotropic Eliashberg spectral function~\cite{Ponce2016}
\begin{equation}
\alpha^{2}F(\omega)= \frac{1}{2}\sum_{\nu}\int_{BZ}\frac{d{\bf q}}{\Omega_{BZ}}\omega_{{\bf q} \nu}\lambda_{{\bf q}\nu}\delta(\omega - \omega_{{\bf q} \nu}),
\label{eq:alpha2F}
\end{equation}
where the phonon-mode resolved EPC strength $\lambda_{{\bf q}\nu}$ is given by
\begin{align}
\lambda_{{\bf q}\nu} = & \frac{1}{N_F\omega_{{\bf q}\nu}}\sum_{nm}\int_{BZ}\frac{d{\bf k}}{\Omega_{BZ}}|g_{mn,\nu}({\bf k},{\bf q})|^{2} \nonumber \\ & \delta(\varepsilon_{n{\bf k}}-\varepsilon_F)\delta(\varepsilon_{m{\bf k}+{\bf q}}-\varepsilon_F).
\end{align}
Here $\omega_{{\bf q}\nu}$ is the eigenfrequency of phonon mode $\nu$ at ${\bf q}$ and $g_{mn,\nu}({\bf k},{\bf q})$ 
is the EPC matrix element between electronic states ${m{\bf k}}$ and ${n{\bf k}+{\bf q}}$.
The electronic state ${n{\bf k}}$-dependent EPC strength $\lambda_{n{\bf k}}$, given by
\begin{equation}
\lambda_{n{\bf k}} = \sum_{\nu m}\int_{BZ}\frac{d{\bf q}}{\Omega_{BZ}}\frac{|g_{mn,\nu}({\bf k},{\bf q})|^{2}}{\omega_{{\bf q}\nu}} \delta(\varepsilon_{m{\bf k}+{\bf q}}-\varepsilon_F),
\end{equation}
also provides useful information about electron-phonon interaction.
The overall EPC strength $\lambda$ is given by the integral of $\alpha^{2}F(\omega)$ over frequency
\begin{equation}
\lambda=2\int\frac{\alpha^{2}F(\omega)}{\omega}d\omega.
\label{eq:lambda}
\end{equation}
In fact, in the past decades, the $T_c$ was often estimated using
the simplified Allen-Dynes McMillan (ADM) formula~\cite{Ponce2016,Babu2019}
\begin{equation}
k_{B}T_{c}=\frac{\hbar \omega_{ln}}{1.20}\mathrm{exp[}\frac{-1.04(1+\lambda)}{\lambda-\mu^{*}_{c}(1+0.62\lambda)}],
\label{eq:ADM}
\end{equation}
where the  EPC parameter $\lambda$ and logarithmically averaged phonon frequency $\hbar\omega_{ln}$
are taken from {\it ab initio} calculations while the effective screened Coulomb potential $\mu^{*}_c$
is generally treated as an empirical parameter. Eliashberg spectral function $\alpha^{2}F(\omega)$ spectra 
shown in Figs. ~\ref{fig:LaPr-ph-bands} and ~\ref{fig:Nd-ph-bands}
represent the frequency-resolved EPC.  The shape of $\alpha^{2}F(\omega)$ closely follows the PhDOS.
We notice that our calculated phonon dispersion, PhDOS and Eliashberg spectral function $\alpha^{2}F(\omega)$
of NdSrNiO are in good agreement with previous {\it ab initio} DFT calculations~\cite{Nomura2019,Li2024}.

Table II lists the calculated EPC constant $\lambda$ of all three nickelates $Re$SrNiO,
which are about 0.15, being several times smaller than the strong-coupled conventional superconductors
such as Pb (1.12) and Nb (0.82)~\cite{McMillan1968,Carbotte1990}.
This places $Re$SrNiO among the weak phonon-mediated superconductors.
Indeed, the calculated $T_c$ of purely EPC-induced superconductivity in $Re$SrNiO
are only about 1.0 K (see Table II). Furthermore, when the screened Coulomb repulsion is switched-on,
the $T_c$ would be reduced to the order of magnitude of 0.01 K (Table II).

\begin{figure*}
\centering
\includegraphics[width=108mm]{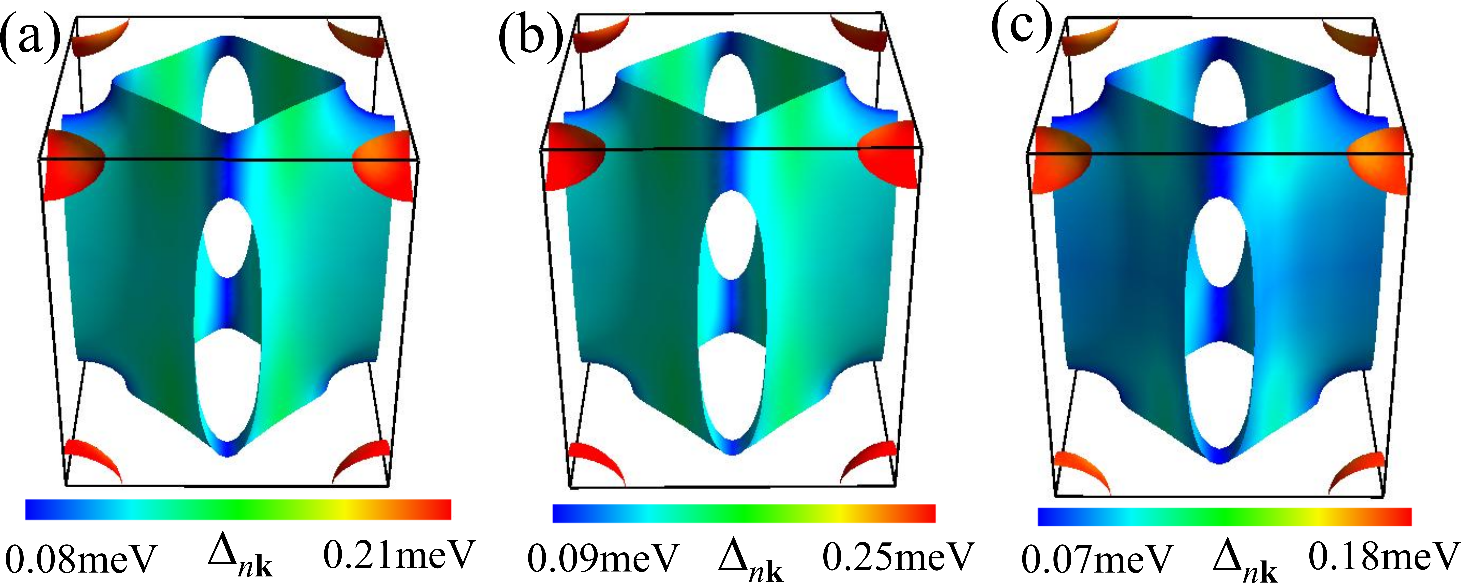}
\caption{Electronic state $n{\bf k}$-dependent purely phonon-mediated superconducting gap $\Delta_{n{\bf k}}$ 
on the Fermi surface of (a) LaSrNiO,  (b) PrSrNiO and (c) NdSrNiO at $T = 0.005$ K. }
\label{fig:EPC_gap_FS}
\end{figure*}

Electronic state $n{\bf k}$-resolved phonon-mediated SC gap functions $\Delta_{n{\bf k}}$ on the FS
in $Re$SrNiO are depicted in Fig. ~\ref{fig:EPC_gap_FS}. First, it is clear from Fig. ~\ref{fig:EPC_gap_FS} 
that all three SC gap functions $\Delta_{n{\bf k}}$ are positive, although their magnitudes 
vary from $\sim$0.07 meV to $\sim$0.25 meV.
In particular, the electron (E-band) FS pockets at the BZ corners have nearly twice larger gap sizes than
the large hole-like (H-band) FS sheet at the BZ center. In other words,
without including the SF interaction,
the three nickelates would be a multiband anisotropic $s$-wave superconductor. This would be similar to
the case of $\gamma$-BiPd, which is a multi-orbital anisotropic $s$-wave superconductor~\cite{Keshri2025}.
Second, electronic state $n{\bf k}$-dependent EPC strengths $\lambda_{n{\bf k}}$ on the FS
are displayed in Fig. ~\ref{fig:lambda_FS} for all three nickelates $Re$SrNiO.
A comparison of Fig. ~\ref{fig:EPC_gap_FS} and Fig. ~\ref{fig:lambda_FS}
shows that for each nickelate, $\lambda_{n{\bf k}}$ and $\Delta_{n{\bf k}}$ have nearly the same distribution
profile on the FS. This should not be surprising. For the purely phonon-mediated superconductivity,
the larger the $\lambda_{n\bf{k}}$ is, the larger the $\Delta_{n\bf{k}}$ will be.

\section{Electronic state-resolved screened Coulomb repulsion and spin fluctuation interaction on the Fermi surface}
\label{sec:appenix_C}
\begin{figure*}
\centering
\includegraphics[width=140mm]{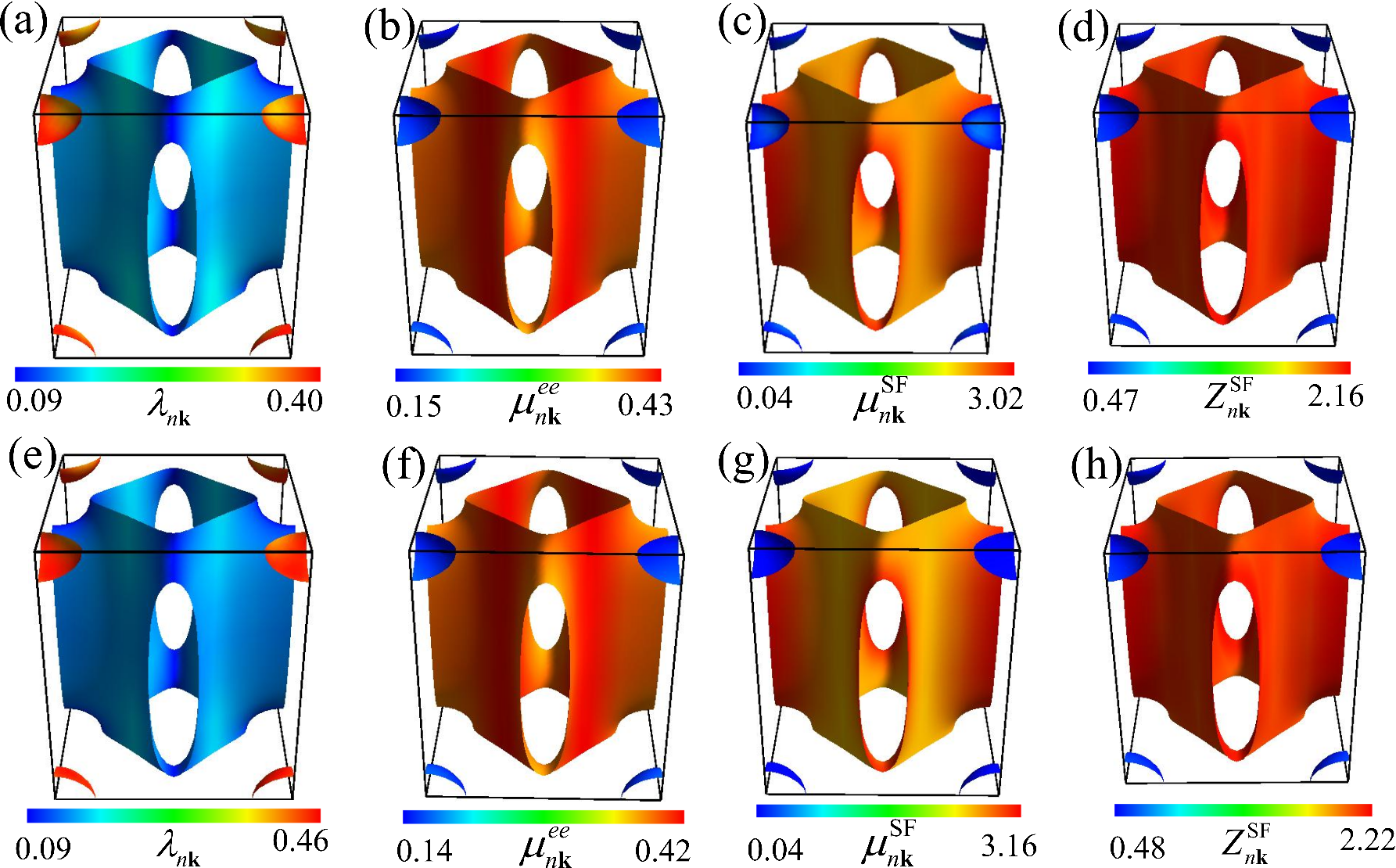}
\includegraphics[width=140mm]{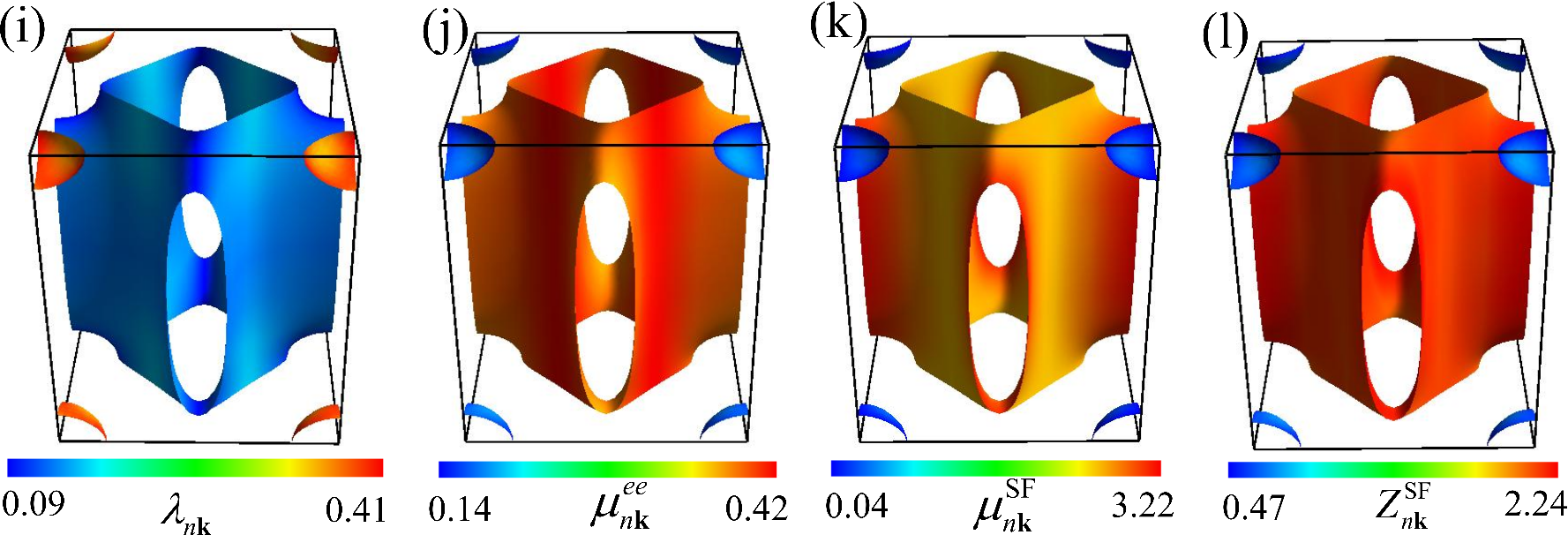}
\caption{Band $n$ and momentum ${\bf k}$-dependent (a,e,i) electron-phonon coupling (EPC) strength $\lambda_{n\bf{k}}$,
(b,f,j) screened Coulomb repulsion $\mu_{n\bf{k}}^{ee}$, (c,g,k) spin-fluctuation (SF) pair-interaction 
$\mu_{n\bf{k}}^{SF}$ and (d,h,l) SF induced renormalization $Z_{n\bf{k}}^{SF}$ on the Fermi surface (FS) 
of (a,b,c,d) LaSrNiO, (e,f,g,h) PrSrNiO and (i,j,k,l) NdSrNiO.
Note that the distribution of EPC induced renormalization $Z_{n\bf{k}}^{ep}$ on the FS
is identical to that of $\lambda_{n\bf{k}}$ (a, e, i), and thus is not shown here.
}
\label{fig:lambda_FS}
\end{figure*}

To gain detailed microscopic insight into the SC pairing mechanism and the SC gap structure,
we present band $n$ and momentum ${\bf k}$-dependent EPC strength $\lambda_{n\bf{k}}$ [see Eq. (B3) in Appendix B],
screened Coulomb repulsion
\begin{equation}
\mu_{n\bf{k}}^{ee}=\sum_{m{\bf k}'}\delta(\xi_{m{\bf k}'})K_{n{\bf k}m{\bf k}'}^{ee}(\xi_{n{\bf k}},\xi_{m{\bf k}'})
\end{equation}
and SF-induced pairing interaction
\begin{equation}
\mu_{n\bf{k}}^{SF}=\sum_{m{\bf k}'}\delta(\xi_{m{\bf k}'})K_{n{\bf k}m{\bf k}'}^{SF}(\xi_{n{\bf k}},\xi_{m{\bf k}'})
\end{equation}
in Fig. ~\ref{fig:lambda_FS} for all three nickelates $Re$SrNiO.
First, Fig. ~\ref{fig:lambda_FS} indicates that on the large cylindrical FS sheet (H-band), $\mu_{n{\bf k}}^{ee}$ 
is always larger than  $\lambda_{n{\bf k}}$. In contrast, $\lambda_{n{\bf k}}$ is larger than $\mu_{n{\bf k}}^{ee}$ 
on the small (E-band) electron pockets at the BZ corners (A points). Since the signs of
the Coulomb repulsion ($\mu_{n{\bf k}}^{ee}$) and phonon-mediated
attraction ($\lambda_{n{\bf k}}$) are opposite, the screened Coulomb repulsion suppresses 
the phonon-mediated attraction on the large H-band FS sheet, while the phonon-mediated attraction 
dominates on the E-band FS pocket.
Since the surface area of the H-band sheet is much larger than that of the E-band pocket,
the large H-band FS sheet plays a dominating role.
This explains why the weak phonon-induced superconductivity in $Re$SrNiO is suppressed when the screened 
e-e Coulomb repulsion is switched-on, as reported in Sec. VI and also can be seen in Table II.
Second, Fig. ~\ref{fig:lambda_FS} shows that the $n{\bf k}$-dependent SF-induced pairing 
interaction $\mu_{n{\bf k}}^{SF}$ is much stronger than both $\mu_{n{\bf k}}^{ee}$ and $\lambda_{n{\bf k}}$
on the large H-band FS sheet. This results in the dominating SF-mediated e-e pairing and hence
rather high $T_c$ superconductivity in $Re$SrNiO.

We also present the $n{\bf k}$-dependent SF interaction ($Z_{n{\bf k}}^{SF}$)
contribution to the renormalization $Z_{n{\bf k}}$ in Fig. ~\ref{fig:lambda_FS}.
Note that the distribution of the $n{\bf k}$-dependent EPC ($Z_{n\bf{k}}^{ep}$) contribution to
the renormalization $Z_{n{\bf k}}$ in $Re$SrNiO is identical to that of $\lambda_{n{\bf k}}$,
and thus is not shown here.
$Z_{n{\bf k}}^{ep}$ and $Z_{n{\bf k}}^{SF}$ represent, respectively, $n{\bf k}$-dependent EPC
and SF-interaction contributions to the renormalization of electronic state $n{\bf k}$.
It is clear from Fig. ~\ref{fig:lambda_FS} that in $Re$SrNiO,
$Z_{n{\bf k}}^{SF}$ (of $\sim$2.0) is much larger than $Z_{n{\bf k}}^{ep}$ ($\lambda_{n\bf{k}}$)
(of $\sim$0.4) on the large H-band FS sheet while they are all small on the E-band FS pocket.
This explains that in LaSrNiO, the calculated H-band near the $E_F$ is strongly
renormalized (by up to 2$\sim$3) and the calculated E-band remains almost unchanged, as revealed
in recent ARPES experiments~\cite{Sun2025}. Nevertheless, as mentioned in Sec. IV,
our calculated Fermi surface in LaSrNiO agrees well with the one observed
in the ARPES experiments~\cite{Sun2025}.

\begin{thebibliography}{}

\bibitem{Bednorz1986} J. G. Bednorz and K. A. M\"{u}ller, Possible high T$_c$ superconductivity
in the Ba-La-Cu-O system, Z. Phys. B {\bf 64}, 908 (1986).

\bibitem{Wu1987} M. K. Wu, J. R. Ashburn, C. J. Torng, P. H. Hor, R. L. Meng, L. Gao, Z. J. Huang, Y. Q. Wang and C. W. Chu, Superconductivity at 93 K in a new mixed-phase Y-Ba-Cu-O compound
system at ambient pressure, Phys. Rev. Lett. {\bf 58}, 908 (1987).

\bibitem{Schilling1993} A. Schilling, M. Cantoni, J. D. Guo and H. R. Ott,
Superconductivity above 130 K in the Hg-Ba-Ca-Cu-O system, Nature (London) {\bf 363}, 56 (1993).

\bibitem{Bardeen1957} J. Bardeen, L. N. Cooper and J. R. Schrieffer, Theory of superconductivity, Phys. Rev. {\bf 108}, 1175 (1957).


\bibitem{Scalapino1999} D. J. Scalapino, Superconductivity and spin fluctuations,
J. Low Temp. Phys. {\bf 117}, 179 (1999).

\bibitem{Tsuei2000} C. C. Tsuei and J. R. Kirtley, Pairing symmetry in cuprate superconductors,
Rev. Mod. Phys. {\bf 72}, 969 (2000).

\bibitem{Hashimoto2014} M. Hashimoto, I. M. Vishik, R.-H. He, T. P. Devereaux and Z.-X. Shen,
Energy gaps in high-transition-temperature cuprate superconductors,
Nature Phys. {\bf 14}, 483 (2014).

\bibitem{Keimer2015} B. Keimer, S. A. Kivelson, M. R. Norman, S. Uchida, and J. Zaanen, 
From quantum matter to high-temperature superconductivity in copper oxides, 
Nature (London) {\bf 518}, 179 (2015). 


\bibitem{Luo2023} X. Luo, H. Chen, Y. Li, Q. Gao, C. Yin, H. Yan, T. Miao, H. Luo, Y. Shu,
Y. Chen {\it et al.}, Electronic origin of high superconducting critical temperature in trilayer cuprates,
Nature Phys. {\bf 19}, 1841 (2023).

\bibitem{Wen2025} C. Wen, Z. Hou, A. Akban, K. Chen, W. Hong, H. Yang, I. Eremin, Y. Li and Hai-Hu Wen,
Unprecedentedly large gap in HgBa$_2$Ca$_2$Cu$_3$O$_{8+\delta}$ with the highest $T_c$ at ambient pressure,
npj Quant. Mater. {\bf 10}, 20 (2025).

\bibitem{Guo1988} G. Y. Guo and W. M. Temmerman, Electronic structure and magnetism in La$_2$NiO$_4$,
J. Phys. C: Solid State Phys. {\bf 21}, L803 (1988).

\bibitem{Anisimov1999} V. I. Anisimov, D. Bukhvalov and T. M. Rice,
Electronic structure of possible nickelate analogs to the cuprates,
Phys. Rev. B {\bf 59}, 7901 (1999).

\bibitem{Lee2004} K.-W. Lee and W. E. Pickett, Infinite-layer nickelate LaNiO$_2$: Ni$^{1+}$ is not Cu$^{2+}$,
Phys. Rev. B {\bf 70}, 165109 (2004).

\bibitem{Hansmann2009} P. Hansmann, X. Yang, A. Toschi, G. Khaliullin, O. K. Andersen and K. Held, 
Turning a nickelate Fermi surface into a cuprate-like one
through heterostructuring, Phys. Rev. Lett. {\bf 103}, 016401 (2009).

\bibitem{Li2019} D. Li, B. Y. Wang, K. Lee, S. P. Harvey, M. Osada, B. H. Goodge, L. F. Kourkoutis, and H. Y. Hwang,
Superconductivity in an infinite-layer nickelate, Nature (London) {\bf 572}, 624 (2019) 

\bibitem{Zeng2020} S. Zeng et al., Phase diagram and superconducting dome of infinite-Layer Nd$_{1-x}$Sr$_x$NiO$_2$ thin films, 
Phys. Rev. Lett.  {\bf 125}, 147003 (2020).

\bibitem{Osada2020} M. Osada, B. Y. Wang, K. Lee, D. Li and H. Y. Hwang,
Phase diagram of infinite layer praseodymium nickelate Pr$_{1-x}$Sr$_x$NiO$_2$ thin films,
Phys. Rev. Mater. {\bf 4}, 121801(R) (2020).

\bibitem{Li2020} D. Li, B. Y. Wang, K. Lee, S. P. Harvey, M. Osada, B. H. Goodge, L. F. Kourkoutis, and H. Y. Hwang,
Superconducting dome in Nd$_{1-x}$Sr$_x$NiO$_2$ infinite layer films, Phys. Rev. Lett. {\bf 125}, 027001 (2020)

\bibitem{Osada2021} M. Osada, B. Y.Wang, B. H. Goodge, S. P. Harvey, K. Lee, D. Li, L. F. Kourkoutis, and H. Y. Hwang, 
Nickelate superconductivity without rare-earth magnetism: (La,Sr)NiO$_2$, Adv. Mater. {\bf 33}, 2104083 (2021).

\bibitem{Lee2023} K. Lee,  B. Y. Wang, M. Osada, B. H. Goodge, T. C. Wang, Y. Lee, S. Harvey, W. J. Kim, Y. Yu
C. Murthy, S. Raghu, L. F. Kourkoutis and H. Y. Hwang, Linear-in-temperature resistivity for optimally superconducting
 (NdSr)NiO$_2$, Nature (London) {\bf 619}, 288 (2023)

\bibitem{Osada2023} M. Osada, K. Fujiwara, T. Nojima and A. Tsukazaki,
Improvement of superconducting properties in La$_{1-x}$Sr$_x$NiO$_2$ thin films by tuning topochemical reduction temperature,
Phys. Rev. Mater. {\bf 7}, L051801 (2023).

\bibitem{Chow2025} S. L. E. Chow, Z. Luo and A. Ariando, Bulk superconductivity near 40 K in hole-doped SmNiO$_2$ at ambient pressure,
Nature (London) {\bf 642}, 58 (2025).

\bibitem{Nomura2019} Y. Nomura, M. Hirayama, T. Tadano, Y. Yoshimoto, K.  Nakamura, and R. Arita, Formation of a two-dimensional
single-component correlated electron system and band engineering in the nickelate superconductor NdNiO$_2$,
Phys. Rev. B {\bf 100}, 205138 (2019).

\bibitem{Botana2020} A. S. Botana and M. R. Norman, Similarities and differences between
LaNiO$_2$ and CaCuO$_2$ and implications for superconductivity, Phys. Rev. X {\bf 10}, 011024 (2020).

\bibitem{Nomura2022} Y. Nomura and R. Arita, Superconductivity in infinitelayer nickelates, Rep. Prog. Phys. {\bf 85}, 052501 (2022).

\bibitem{Wu2020} X. Wu, D. D. Sante, T. Schwemmer, W. Hanke, H. Y. Hwang, S. Raghu and R. Thomale,
Robust $d_{x^2-y^2}$-wave superconductivity of infinite-layer nickelates, Phys. Rev. B \textbf{101}, 060504 (2020).

\bibitem{Sakakibara2020} H. Sakakibara, H. Usui, K. Suzuki, T. Kotani, H. Aoki, and K. Kuroki,
Model construction and a possibility of cupratelike pairing in a new $d^9$ nickelate superconductor (Nd,Sr)NiO$_2$,
Phys. Rev. Lett. {\bf 125}, 077003 (2020)

\bibitem{Harvey2025} S. P. Harvey, B. Y. Wang, J. Fowlie, M. Osada, K. Lee, Y. Lee, D. Li, and H. Y. Hwang,
Evidence for nodal superconductivity in infinite-layer nickelates, PNAS {\bf 122}, e2427243122 (2025).

\bibitem{Gu2020} Q. Gu, Y. Li, S. Wan, H. Li, W. Guo, H. Yang, Q. Li, X. Zhu, X. Pan, Y. Nie and H.-H. Wen,
Single particle tunneling spectrum of superconducting Nd$_{1-x}$Sr$_x$NiO$_2$ thin films,
Nature Commun. {\bf 11}, 6027 (2020)

\bibitem{Chow2022} L. E. Chow,  S. K. Sudheesh, Z. Y. Luo, P. Nandi, T. Heil, J. Deuschle, S. W. Zeng, Z. T. Zhang, S. Prakash, X. M. Du, Z. S. Lim, P. A. van Aken, E. E. M. Chia, A. Ariando, Pairing symmetry in infinite-layer nickelate superconductor, arXiv:2201.10038v2

\bibitem{Li2024} Z. Li and S. G. Louie, Two-Gap Superconductivity and the Decisive Role of Rare-Earth $d$ Electrons
in Infinite-Layer Nickelates, Phys. Rev. Lett. \textbf{133}, 126401 (2024).

\bibitem{Oliveira1988} L. N. Oliveira, E. K. U. Gross and W. Kohn, Density-functional theory for superconductors,
Phys. Rev.  Lett. {\bf 60}, 2430 (1988).

\bibitem{Lueders2005} M. Lueders, M. A. L. Marques, N. N. Lathiotakis, A. Floris, G. Profeta, L. Fast,
A. Continenza, S. Massidda and E. K. U. Gross, {\it Ab initio} theory of superconductivity.
I. Density functional formalism and approximate functionals, Phys. Rev.  B {\bf 72}, 024545 (2005).

\bibitem{Marques2005} M. A. L. Marques, M. Lueders, N. N. Lathiotakis, G. Profeta, A. Floris, L. Fast,
A. Continenza and E. K. U. Gross, {\it Ab initio} theory of superconductivity.
II. Application to elemental metals, Phys. Rev.  B {\bf 72}, 024545 (2005).

\bibitem{Kawamura2020} M. Kawamura, Y. Hizume and T. Ozaki, Benchmark of density functional theory
for superconductors in elemental materials, Phys. Rev.  B {\bf 101}, 134511 (2020).

\bibitem{Akashi2013} R. Akashi and R. Arita, Development of density-functional theory for
a plasmon-assisted superconducting state: Application to lithium under higher pressure,
Phys. Rev.  Lett. {\bf 111}, 057006 (2013).

\bibitem{Essenberger2014} F. Essenberger, A. Sanna, A. Linscheid, F. Tandetzky, G. Profeta, P. Cudazzo
and E. K. U. Gross, Superconducting pairing mediated by spin fluctuations from first principles
 Phys. Rev.  B {\bf 90}, 214504 (2014).

\bibitem{Floris2005} A. Floris, G. Profeta, N. N. Lathiotakis, M. L\"{u}ders, M. A. L. Marques, C. Franchini and E. K. U. Gross,
Superconducting properties of MgB$_2$ from first principles,
Phys. Rev.  Lett. {\bf 94}, 037004 (2005).

\bibitem{Baroni2001} S. Baroni, S. de Gironcoli, A. Dal Corso, and P. Giannozzi,
\href{https://link.aps.org/doi/10.1103/RevModPhys.73.515}{Phonons and related crystal properties from density-functional perturbation theory},
Rev. Mod. Phys. \textbf{73}, 515 (2001).

\bibitem{Gell-Mann1957} M. Gell-Mann and K. Brueckner, Correlation Energy of an Electron Gas at High Density,
Phys. Rev.  {\bf 106}, 364 (1957).

\bibitem{Zangwill1980} A. Zangwill and P. Soven, Density-functional approach to local-field effects in finite systems: Photoabsorption in the rare gases, Phys. Rev.  A {\bf 21}, 1561 (1980).

\bibitem{Tsutsumi2020} K. Tsutsumi, Y. Hizume, M. Kawamura, R. Akashi and S. Tsuneyuki,
Effect of spin fluctuations on superconductivity in V and Nb: A first-principles study,
Phys. Rev.  B {\bf 102}, 214515 (2020).

\bibitem{Hayward2003} M. A. Hayward and M. J. Rosseinsky, Synthesis of the infinite layer Ni(I) phase
NdNiO$_{2+x}$ by low temperature reduction of NdNiO$_3$ with sodium hydride, Solid State Sci. {\bf 5}, 839 (2023).


\bibitem{Sun2025} W. Sun, Z. Jiang, C. Xia, B. Hao, S. Yan, M. Wang, Y. Li, H. Liu, J. Ding, J. Liu, Z. Liu, J. Liu, H. Chen, D. Shen and Y. Nie,
Electronic structure of superconducting infinite-layer lanthanum nickelates,
Sci. Adv. {\bf 11}, eadr5116 (2025)

\bibitem{Perdew1981} J. P. Perdew and A. Zunger, Self-interaction correction to density-functional approximations
for many-electron systems, Phys. Rev. B {\bf 23}, 5048 (1981).

\bibitem{Corso2014} A. Dal Corso, Pseudopotentials periodic table: From H to Pu, Comp. Mater. Sci. {\bf 95}, 337 (2014).

\bibitem{Pslibrary} https://www.quantum-espresso.org/pseudopotentials.

\bibitem{Giannozzi2009} P. Giannozzi, S. Baroni, N. Bonini, M. Calandra, R. Car,
C. Cavazzoni, D. Ceresoli, G. L. Chiarotti, M. Cococcioni, I. Dabo, $et$ $al$.,
\href{https://doi.org/10.1088/0953-8984/21/39/395502}{QUANTUM ESPRESSO: a modular and open-source software project for quantum simulations of materials},
J. Phys.: Condens. Matter \textbf{21}, 395502 (2009).

\bibitem{Giannozzi2017} P. Giannozzi, O. Andreussi, T. Brumme, O. Bunau,
M. B. Nardelli, M. Calandra, R. Car, C. Cavazzoni, D. Ceresoli, M. Cococcioni, $et$ $al$.,
\href{https://doi.org/10.1088/1361-648X/aa8f79}{Advanced capabilities for materials modelling with Quantum ESPRESSO},
J. Phys.: Condens. Matter \textbf{29}, 465901 (2017).

\bibitem{Kawamura2014} M. Kawamura, Y. Gohda and S. Tsuneyuki, Improved tetrahedron method for the Brillouin-zone
integration applicable to response function, Phys. Rev.  B {\bf 89}, 094515 (2014).

\bibitem{sctk} http://sctk.osdn.jp/

\bibitem{Kawamura2019} M. Kawamura, FermiSurfer: Fermi-surface viewer providing multiple representation schemes,
 Comp. Phys. Commun. {\bf 239}, 197 (2019)

\bibitem{Sakakibara2012} H. Sakakibara, H. Usui, K. Kuroki, R. Arita and H. Aoki,
Origin of the material dependence of $T_c$ in the single-layered cuprates,
Phys. Rev. B {\bf 85}, 064501 (2012).

\bibitem{Annett1990} J. F. Annett, Symmetry of the order parameter for high-temperature superconductivity,
Adv. Phys. {\bf 39}, 83 (1990).

\bibitem{Yip1993} S. Yip and Anupam Garg, Superconducting states of reduced symmetry: General order parameters
and physical implications, Phys. Rev. B {\bf 48}, 3304 (1993).

\bibitem{Carbotte1990} J. P. Carbotte, Properties of boson-exchange superconductors, Rev. Mod. Phys. {\bf 62}, 1027 (1990).

\bibitem{Sigrist2005} M. Sigrist, Introduction to unconventional superconductivity, AIP Conf. Proc. {\bf 789}, 165 (2005).

\bibitem{Keshri2025} S. P. Keshri and G.-Y. Guo, {\it Ab initio} study of orbital-selective superconductivity in $\gamma$-BiPd,
Phys. Rev.  B {\bf 112}, 214518 (2025).

\bibitem{Janak1977} J. F. Janak, Uniform susceptibilities of metallic elements, Phys. Rev. B {\bf 16}, 255 (1977).

\bibitem{Kawamura2017} M. Kawamura, R. Akashi and S. Tsuneyuki, Anisotropic superconducting gaps in YNi$_2$B$_2$C:
A first-principles investigation, Phys. Rev. B {\bf 95}, 054506 (2017).

\bibitem{Prozorov2006} R. Prozorov and R.W. Giannetta, Magnetic penetration depth in unconventional superconductors,
Supercond. Sci. Technol. {\bf 19}, R41 (2006).

\bibitem{Cheng2024} B. Cheng, D. Cheng, K. Lee, L. Luo, Z. Chen, Y. Lee, B. Y. Wang, M. Mootz, I. E. Perakis, Z.-X.
Shen, H. Y. Hwang, and J. Wang, Evidence for d-wave superconductivity of infinite-layer nickelates
from low-energy electrodynamics, Nat. Mater. {\bf 23}, 775 (2024).












\bibitem{Ponce2016} S. Ponce, E. R. Margine, C. Verdi, and F. Giustino,
\href{https://doi.org/10.1016/j.cpc.2016.07.028}{EPW: Electron--phonon coupling, transport and superconducting 
properties using maximally localized Wannier functions},
Computer Physics Communications \textbf{209}, 116 (2016).

\bibitem{Babu2019} K. R. Babu and G.-Y. Guo, Electron-phonon coupling, superconductivity, and nontrivial band topology in NbN polytypes,
Phys. Rev. B {\bf 99}, 104508 (2019).

\bibitem{McMillan1968} W. L. McMillan,
\href{https://link.aps.org/doi/10.1103/PhysRev.167.331}{Transition Temperature of Strong-Coupled Superconductors},
Phys. Rev. \textbf{167}, 331 (1968).

\end{thebibliography}

\end{document}